\documentclass[a4paper,8pt]{article}
\pdfoutput=1

\usepackage{fancyhdr}

\usepackage[font=small,labelfont=bf]{caption}
\DeclareCaptionFont{scriptsize}{\scriptsize}
\DeclareCaptionFont{tiny}{\tiny}

\usepackage{float}
\usepackage{amsmath, amssymb, setspace,cite,verbatim}
\usepackage{hyperref}
\usepackage{color}
\usepackage{wrapfig}

\usepackage{multicol}
\usepackage{etoolbox}
\usepackage{relsize}
\usepackage{multirow}

\usepackage{latexsym}
\usepackage{a4wide}
\usepackage{graphicx, subfigure}
\usepackage{tabularx}
\usepackage{array}
\usepackage{graphics}     
\usepackage{longtable} 
\usepackage[table]{xcolor}
\usepackage{booktabs}
\usepackage[utf8]{inputenc}
\usepackage{slashed}

\usepackage{pifont}

\definecolor{light-gray}{gray}{0.90}

\renewcommand{\arraystretch}{1.2} 
 
\newcommand{\cen}[1]{\multicolumn{1}{c}{#1}}
\newcommand{\ra}[1]{\renewcommand{\arraystretch}{#1}}
\newcommand{\rb}[1]{\renewcommand{\tabcolsep}{#1}}

\begin{document}

\hfill {\tt  CERN-TH-2018-134, IPM/P.A-507, MITP/18-046}  

\def\thefootnote{\fnsymbol{footnote}}
 
\begin{center}

\vspace{3.cm}

{\Large\bf Hadronic and New Physics Contributions to $b \to s$ Transitions}

\setlength{\textwidth}{11cm}
                    
\vspace{2.cm}
{\large\bf  
A.~Arbey$^{\,a,b,}$\footnote{Also Institut Universitaire de France, 103 boulevard Saint-Michel, 75005 Paris, France}$^{,}$\footnote{Email: alexandre.arbey@ens-lyon.fr}, 
T.~Hurth$^{c,}$\footnote{Email: tobias.hurth@cern.ch},
F.~Mahmoudi$^{a,b,*,}$\footnote{Email: nazila@cern.ch},
S.~Neshatpour$^{d,}$\footnote{Email: neshatpour@ipm.ir }
}
 
\vspace{1.cm}
{\em $^a$Univ Lyon, Univ Lyon 1, CNRS/IN2P3, Institut de Physique Nucl\'eaire de Lyon, UMR5822,\\ F-69622 Villeurbanne, France}\\[0.2cm]
{\em $^b$Theoretical Physics Department, CERN, CH-1211 Geneva 23, Switzerland} \\[0.2cm]
{\em $^c$PRISMA Cluster of Excellence and  Institute for Physics (THEP)\\
Johannes Gutenberg University, D-55099 Mainz, Germany}\\[0.2cm] 
{\em $^d$School of Particles and Accelerators,
Institute for Research in Fundamental Sciences (IPM)\\
P.O. Box 19395-5531, Tehran, Iran}

\end{center}

\renewcommand{\thefootnote}{\arabic{footnote}}
\setcounter{footnote}{0}

\vspace{1.cm}
\thispagestyle{empty}
\centerline{\bf ABSTRACT}
\vspace{0.5cm}
Assuming the source of the anomalies observed recently in $b \to s$ data to be new physics, there is a priori no reason to believe that -- in the effective field theory language -- only one type of operator is responsible for the tensions. We thus perform  for the first time a global fit where all the Wilson coefficients which can effectively receive new physics contributions are considered, allowing for lepton flavour universality breaking effects as well as contributions from chirality flipped and scalar and pseudoscalar operators, and find the SM pull taking into account all effective parameters. 
As a result of the full fit to all available $b \to s$ data including all relevant Wilson coefficients, we obtain a total pull of 4.1$\sigma$ with the SM hypothesis assuming 10\% error for the power corrections.
Moreover, we make a statistical comparison to find whether the most favoured explanation of the anomalies is new physics or underestimated hadronic effects using the most general parameterisation which is fully consistent with the analyticity structure of the amplitudes. This Wilks' test will be a very useful tool to analyse the forthcoming $B\to K^* \mu^+ \mu^-$ data.
Because the significance of the observed tensions in the angular observables in $B \to K^* \mu^+\mu^-$  is presently dependent on the theory estimation of the hadronic contributions to these decays, we briefly discuss the various available approaches for taking into account the long-distance hadronic effects and examine how the different estimations of these contributions result in distinct significance of the new physics interpretation of the observed anomalies.

\newpage
\section{Introduction}
Currently among the most significant particle physics measurements hinting to the observation of new physics  (NP) are the  tensions between the Standard Model (SM) predictions and the corresponding experimental measurements in several $b \to s \ell^+ \ell^-$ decays.
The first tension was observed in the angular observable $P_5^\prime$ in the $B\to K^* \mu^+ \mu^-$ decay with 1 fb$^{-1}$  of data~\cite{Aaij:2013qta} at the LHCb experiment with a significance of more than $3\sigma$ and later confirmed by the same experiment with 3 fb$^{-1}$ of data~\cite{Aaij:2015oid}.
$B\to K^* \mu^+ \mu^-$ angular observables were also measured by the Belle~\cite{Abdesselam:2016llu}, ATLAS~\cite{Aaboud:2018krd} and CMS~\cite{CMS-PAS-BPH-15-008} experiments with larger experimental uncertainties.
Another measurement indicating larger than $3\sigma$ tension with the SM was performed by the LHCb~\cite{Aaij:2015esa} in the branching ratio of $B_s \to \phi \mu^+ \mu^-$. 
Several other tensions with a NP significance of $2.2-2.6\sigma$ have also been measured in the ratios $R_K$ and $R_{K^{*}}$ by the LHCb~\cite{Aaij:2014ora,Aaij:2017vbb}. These tensions in the ratios if confirmed would establish the breaking of lepton flavour universality. 
Moreover, smaller tensions with the SM predictions (between 1 and 3$\sigma$) are observed in the branching ratios of $B^0\to K^0 \mu^+ \mu^-$, $B^+\to K^+ \mu^+ \mu^-$, $B^+\to K^{*+} \mu^+ \mu^-$~\cite{Aaij:2014pli} as well as in the baryonic decay of $\Lambda_b\to \Lambda \mu^+ \mu^-$~\cite{Aaij:2015xza}. All the tensions observed  are in the decays with muons in the final state, at low dilepton invariant mass squared ($q^2$) and the measurements are below the SM predictions. The tensions in the branching ratios, angular observables or $R$ ratios point to a coherent picture of deviations with the SM and they can all be explained with a common NP effect, namely about 25\% reduction in the $C_9^{(\mu)}$ Wilson coefficient~\cite{Hurth:2014vma,Hurth:2013ssa,Hurth:2017hxg} (see also Refs.~\cite{Capdevila:2017bsm,Altmannshofer:2017yso,DAmico:2017mtc,Hiller:2017bzc,Geng:2017svp,Ciuchini:2017mik}).

While the tensions in the ratios are not very significant and below 3$\sigma$ at the moment, in case they are confirmed by further experimental data, the only viable explanation would be NP since the theory predictions of these observables are very precise and robust~\cite{Hiller:2003js,Bordone:2016gaq} due to hadronic cancellations.
On the other hand, the observables $P_5^\prime(B\to K^* \mu^+ \mu^-)$ and BR($B_s \to \phi \mu^+ \mu^-$) both receive hadronic contributions which are difficult to estimate, especially the ones emerging from non-factorisable power corrections. 
Nevertheless, the confirmation of the $P_5^\prime(B\to K^* \mu^+ \mu^-)$ anomaly by several measurements makes it unlikely that the tension in $P_5^\prime$ is due to statistical fluctuations and hence either underestimated hadronic effects or NP contributions are the more likely explanations~\cite{Jager:2012uw,Jager:2014rwa,Descotes-Genon:2013wba,Altmannshofer:2013foa,Hambrock:2013zya,Beaujean:2013soa,Horgan:2013pva,Hurth:2013ssa,Mahmoudi:2014mja,Descotes-Genon:2014uoa,Khodjamirian:2010vf,Khodjamirian:2012rm,Lyon:2014hpa,Descotes-Genon:2015uva}.
The significance of the tension in $P_5^\prime$ depends on the precise treatment of the hadronic contributions~\cite{Hurth:2013ssa,Hurth:2016fbr,Mahmoudi:2016mgr}.

The angular observables of the $B\to K^* \ell^+ \ell^-$ decay can be constructed in such a way to minimise the hadronic uncertainties emerging from form factor contributions~\cite{Egede:2008uy,Egede:2010zc}. While an appropriate choice could offer specific form factor independent observables (at leading order), when considering the full set of angular observables and taking into account the correlations (both experimental and theoretical) in the global fit, the uncertainty merely shifts from one observable to another and a change of basis would not offer further physical information~\cite{Hurth:2016fbr}.
Moreover, another source of hadronic uncertainties is due to non-local contributions from four-quark 
operators, especially from charm loops, which give rise to the non-factorisable power corrections.

The standard framework for the  calculation of the non-factorisable hadronic contributions in the $B\to K^* \ell^+ \ell^-$ decay~\cite{Kruger:2005ep,Bobeth:2008ij,Egede:2008uy,Altmannshofer:2008dz,Egede:2010zc}, in the region where $q^2$ is below the $J/\psi$ resonance, is the QCD factorisation (QCDf) method where an expansion of $\Lambda/m_b$ is employed~\cite{Beneke:2001at,Beneke:2004dp}.  
Within this framework higher powers of $\Lambda/m_b$ remain unknown and are usually roughly estimated to be some fraction of the known leading order QCDf terms.
However, there have been methods suggested for the estimation of the power corrections using light-cone sum rule (LCSR) techniques and employing dispersion relations~\cite{Khodjamirian:2010vf}
and the analyticity structure of the amplitudes~\cite{Bobeth:2017vxj} as well as an empirical model where the hadronic resonances are described as Breit–Wigner amplitudes~\cite{Blake:2017fyh}.
In this paper we investigate how the different methods impact the $B\to K^* \ell^+ \ell^-$ observables and in particular study the tension with $P_5^\prime$ within the several available implementations of the power corrections. We also examine how the significance of the preferred NP  scenarios changes depending on the employed method for estimating the power corrections.

Alternatively, instead of making assumptions on the size of the power corrections they can be parameterised by a general function with a number of unknown free parameters~\cite{Ciuchini:2015qxb,Chobanova:2017ghn,Neshatpour:2017qvi,Ciuchini:2017mik} and then fitted to the data. In this case it is important to have the correct description of the general function and to avoid disruption of the analyticity structure of the amplitude.
Specifically, the ansatz should be in such a way as not to generate a pole in the longitudinal amplitude of the $B\to K^* \ell^+ \ell^-$ especially if the data to $B \to K^* \gamma$ is to be considered since the longitudinal amplitude should vanish when the intermediate  $\gamma$ becomes on-shell. 
We present here for the first time a statistical comparison of both hadronic parameters and NP  contributions to Wilson coefficients within this general parameterisation using the Wilks' theorem \cite{Wilks:1938dza}.

Considering the $b \to s \ell^+ \ell^-$ anomalies to be due to NP   contributions there is a priori no reason to assume that such contributions only appear in a single operator and in principle several operators could simultaneously affect $b \to s \ell^+ \ell^-$ transitions. 
We discuss how the BR($B_s \to \mu^+ \mu^-$) observable, which is usually used to neglect potential contributions from scalar and pseudoscalar operators, cannot be solely considered for such a conclusion. We also take into account that there are regions of parameter space that allow for large contributions to these operators and that in order to  disregard the  scalar and pseudoscalar contributions, all $b \to s$ transitions  should be globally considered.
We perform NP fits  in the most general case where all the relevant Wilson Coefficients including the scalar and pseudoscalar operators, can receive NP  contributions and explore how well scenarios with extended NP contributions describe the $b \to s$ data. We examine whether indeed simultaneous contributions to several operators are favoured or not. This also allows us for the first time to determine the SM pull taking into account all effective number of degrees of freedom, in which the insensitive coefficients are not counted.

The rest of the paper is organised as follows. In section~\ref{sec:NPvsHad} the general ansatz for the power corrections which respects the analyticity of the amplitude is given 
where we make statistical comparisons of the hadronic and NP fits to $B\to K^* \mu^+ \mu^-$ observables.
In section~\ref{sec:implementations} we discuss the various methods available for implementing the hadronic contributions relevant to $B\to K^* \ell^+ \ell^-$ decay and 
examine the most favoured scenarios and the corresponding significance depending on the employed method.
Finally, we discuss the global fit to all possible Wilson coefficients which impact the $b\to s \ell^+ \ell^-$ transitions including scalar contributions  in section~\ref{sec:allWilson}, and give our conclusions in section~\ref{sec:conclusions}.

\section{Hadronic versus NP contributions in $B\to K^* \ell^+ \ell^-$}\label{sec:NPvsHad}

The $b \to s  \ell^+ \ell^-$ transitions are described via an effective Hamiltonian which can be separated into a hadronic and a semileptonic part:
\begin{equation}
 {\cal H}_{\rm eff} = {\cal H}_{\rm eff}^{\rm had} + {\cal H}_{\rm eff}^{\rm sl}\,,
\end{equation}
where
\begin{align}
 {\cal H}_{\rm eff}^{\rm had}&=-\frac{4G_F}{\sqrt{2}}V_{tb}V_{ts}^*\sum_{i=1,\ldots,6,8}C_i\; O_i\,, \nonumber \\
 \quad {\cal H}_{\rm eff}^{\rm sl}&=-\frac{4G_F}{\sqrt{2}}V_{tb}V_{ts}^*
 \sum_{i=7,9,10,Q_1,Q_2,T}(C_i\; O_i + C^\prime_i\; O^\prime_i)\,.
\end{align}
For the exclusive decays $B\to K^* \mu^+ \mu^-$ and $B_s \to \phi \mu^+ \mu^-$, the semileptonic part of the
Hamiltonian which accounts for the dominant contribution, can be described by seven independent form factors 
$\tilde{S}, \tilde{V}_\lambda, \tilde{T}_\lambda$, with helicities $\lambda=\pm1,0$. 
The exclusive $B\to V \bar \ell \ell$ decay, where $V$ is a vector meson can be described  by the following eight helicity amplitudes:
\begin{align}\label{eq:HV}     
 H_V(\lambda) &=-i\, N^\prime \Big\{ C_9^{\rm eff} \tilde{V}_{\lambda} - C_{9}'  \tilde{V}_{-\lambda}
      + \frac{m_B^2}{q^2} \Big[\frac{2\,\hat m_b}{m_B} (C_{7}^{\rm eff} \tilde{T}_{\lambda} - C_{7}'  \tilde{T}_{-\lambda})
      - 16 \pi^2 {\cal N}_\lambda \Big] \Big\} ,  \\  
  H_A(\lambda) &= -i\, N^\prime (C_{10}  \tilde{V}_{\lambda} - C_{10}'\tilde{V}_{-\lambda}) , \\
  H_P &= i\, N^\prime \left\{  (C_{Q_2} - C_{Q_2}')         
            + \frac{2\,m_\ell \hat m_b}{q^2} \left(1+\frac{m_s}{m_b} \right)(C_{10}-C_{10}') \right\} \tilde{S}, \\
  H_S &= i\, N^\prime (C_{Q_1} - C_{Q_1}')\tilde{S}, 
\end{align}
where the effective part of $C_9^{\rm eff}\left(\equiv C_9+Y(q^2)\right)$ as well as the 
non-factorisable contribution ${\cal N}_\lambda(q^2)$ arise 
from the hadronic part of the Hamiltonian through the emission of a photon which itself turns into a lepton pair.
Due to the vectorial coupling of the photon to the lepton pair, the contributions of ${\cal H}_{\rm eff}^{\rm had}$ appear in the vectorial helicity amplitude $H_V(\lambda)$.
It is due to the similar effect from the short-distance $C_9$ (and $C_7$) of ${\cal H}_{\rm eff}^{\rm sl}$ and the long-distance contribution from ${\cal H}_{\rm eff}^{\rm had}$ that there is an ambiguity in separating NP effects of the type $C_9^{\rm NP}$ (and $C_7^{\rm NP}$) from non-factorisable hadronic contributions.

\subsection{Most general ansatz for the non-factorisable power corrections}\label{sec:hlmabda}

The non-factorisable term ${\cal N}_\lambda(q^2)$ contributing to $H_V(\lambda)$ is known at leading order in $\Lambda/m_b$ from QCDf calculations while higher powers can only be guesstimated within QCDf. 
These power corrections are usually assumed to be 10\%, 20\%, etc. of the leading order non-factorisable contribution.
On the other hand, instead of making such a guesstimate on the size of the power corrections they can be parameterised by a polynomial with a number of free parameters which can be fitted to the experimental data~\cite{Ciuchini:2015qxb}.

In our previous work (Ref.~\cite{Chobanova:2017ghn}) 
we assumed a general $q^2$-polynomial ansatz for the unknown contributions
\begin{align}\label{eq:hlambda}
  h_\lambda(q^2)= h_\lambda^{(0)} + \frac{q^2}{1 {\rm GeV}^2}h_\lambda^{(1)} 
  + \frac{q^4}{1 {\rm GeV}^4}h_\lambda^{(2)}\,. 
\end{align}
We used the measurements on $B \to K^* \mu^+ \mu^-$ observables below the $J/\psi$ resonance to fit the free parameters $h^{(0,1,2)}_\lambda$.
However, it turns out that this ansatz that was used in~\cite{Ciuchini:2015qxb} is not compatible with the general analyticity structure of the amplitude $H_V(\lambda)$ in the case of $\lambda=0$, in particular there should be no physical pole in the longitudinal amplitude for $q^2 \to 0$ which is relevant when the branching ratio of $B\to K^* \gamma$ decay is considered and in principle can affect the results. 
In the current paper, we also consider the experimental result on BR$(B\to K^* \gamma)$, thus compatibility with the analytical structure for $q^2 \to 0$ is mandatory. 

We have therefore modified the $h_\lambda(q^2)$ ansatz for $\lambda=0$ and have kept the same ansatz for $\lambda=\pm$ (see appendix~\ref{sec:q2behaviour})
\begin{align}\label{eq:hzero}
  h_0(q^2)= \sqrt{q^2}\times \left( h_0^{(0)} + \frac{q^2}{1 {\rm GeV}^2}h_0^{(1)} + \frac{q^4}{1 {\rm GeV}^4}h_0^{(2)}\right).
\end{align}
This modified definition for $h_{\lambda}$ is the most general ansatz for the unknown hadronic contributions 
(up to higher order powers in $q^2$) which is compatible with the analyticity structure 
assumed in Ref.~\cite{Bobeth:2017vxj}. 

The radiative decay $B \to K^* \gamma$ can be described in terms of  the helicity amplitudes $H_V(\lambda=\pm)$~\cite{Jager:2012uw}
\begin{align}
{\cal A}_\lambda(\bar B \to \bar{K}^* \gamma) &= \lim_{q^2 \to 0} \frac{q^2}{e} H_V(q^2=0; \lambda) \nonumber \\
    &= \frac{i N m_B^2}{e} \left[\frac{2 \hat m_b}{m_B} (C_7 \tilde T_\lambda(0) - C_7' \tilde T_{-\lambda}(0) - 16 \pi^2 {\cal N}_\lambda(q^2=0) \right] .
\end{align}
with ${\cal N}_\lambda (q^2) \equiv \text{leading order in QCDf} + h_\lambda (q^2=0)$
where the leading order contributions in QCDf include the vertex corrections, 
spectator scattering and weak annihilation contributions and can be found in Refs.~\cite{Beneke:2001at,Asatryan:2001zw,Kagan:2001zk,Feldmann:2002iw,Beneke:2004dp}.
With the description in Eq.~(\ref{eq:hzero}) for the power corrections, the $B \to K^* \gamma$ decay can also be described correctly
without developing a pole at $q^2 \to 0$.

We show in appendix~\ref{sec:q2behaviour} that the effect of NP contributions to $B \to K^* \ell^+ \ell^-$ observables from $C_7$ and $C_9$ 
can be embedded in the most general ansatz of the hadronic contributions. Thus  it is possible to make a statistical comparison of a hadronic fit and a NP fit of $C_9$ (and $C_7$) to the $B\to K^* \mu^+ \mu^-$ data.

\subsection{Hadronic fit vs NP fit to $\delta C_{7,9}$}\label{sec:hadronic_results}
In order to investigate whether the $B \to K^* \mu^+ \mu^-$ data are better explained by assuming NP or underestimated hadronic contributions, 
we have done separate fits for each case where only the low $q^2$  data have been used (see also Ref.~\cite{Chobanova:2017ghn}).
For the fits we have considered BR$(B\to K^* \gamma)$~\cite{Amhis:2014hma}, 
${\rm BR}(B^+ \to K^{+*} \mu^+ \mu^-)_{q^2 \in [1.1-6.0]\;{\rm GeV}^2}$~\cite{Aaij:2014pli} and the CP averaged observables of the $B \to K^* \mu^+ \mu^-$  decays~\cite{Aaij:2015oid,Aaij:2016flj} 
in the low $q^2$ bins up to 8~GeV$^2$.
For the theory predictions {\tt SuperIso v4.0}~\cite{Mahmoudi:2007vz,Mahmoudi:2008tp} has been used. 
The SM prediction of $B^{(+)} \to K^{(+)*} \mu^+ \mu^-$ observables
can be found in Ref.~\cite{Hurth:2016fbr}. Using the ``LCSR+Lattice'' result for the $T_1(0)$ form factor~\cite{Straub:2015ica}  we have  BR$(B \to K^* \gamma)=(4.29 \pm 0.85)\times10^{-5}$ (see e.g. Ref.~\cite{Paul:2016urs} regarding the effect of the form factor choice).

For the hadronic fit, employing the parameterisation of section~\ref{sec:hlmabda}, we have varied the 18 free parameters describing the complex $h_{+,-,0}^{(0,1,2)}$.
Most of the fitted parameters are consistent with zero (see Table~\ref{tab:imp1_set78_Had}) as they have large uncertainties, however, this  
can be changed with more precise experimental results and finer $q^2$ binning in the future.

\begin{table}[t!]
\ra{0.90}
\rb{1.3mm}
\begin{center}
\setlength\extrarowheight{2pt}
\scalebox{0.9}{
\begin{tabular}{|l||r|r|}
\hline
 \multicolumn{3}{|c|}{Observables in the low $q^2$ bins up to 8 GeV$^2$}           \\  
 \multicolumn{3}{|c|}{ ($\chi^2_{\rm SM}=54.9,\; \chi^2_{\rm min}=14.7$)}   \\ 
\hline
& \cen{Real}                         & \multicolumn{1}{c|}{Imaginary}   \\ 
 \hline
$h_+^{(0)}$ & $ ( 1.67 \pm 2.15 ) \times 10^{-4}$     &  $ (-1.17 \pm 1.84 ) \times 10^{-4}$ \\ 
$h_+^{(1)}$ & $ ( 1.55 \pm 32.01 ) \times 10^{-5}$    &  $ (-1.65 \pm 2.35 ) \times 10^{-4}$ \\ 
$h_+^{(2)}$ & $ ( -1.65 \pm 72.01 ) \times 10^{-6}$   &  $ (4.36 \pm 3.73 ) \times 10^{-5}$ \\ 
\hline 
$h_-^{(0)}$ & $ ( -2.13 \pm 1.77 ) \times 10^{-4}$    &  $ (4.79 \pm 3.24 ) \times 10^{-4}$ \\ 
$h_-^{(1)}$ & $ ( 3.69 \pm 12.56 ) \times 10^{-5}$    &  $ (-5.31 \pm 3.71 ) \times 10^{-4}$ \\ 
$h_-^{(2)}$ & $ ( 1.29 \pm 1.84 ) \times 10^{-5}$     &  $ (5.79 \pm 6.93 ) \times 10^{-5}$ \\ 
\hline 
$h_0^{(0)}$ & $ ( -3.61 \pm 36.99 ) \times 10^{-5}$   &  $ (6.89 \pm 4.52 ) \times 10^{-4}$ \\ 
$h_0^{(1)}$ & $ ( 3.63 \pm 2.98 ) \times 10^{-4}$     &  $ (-6.52 \pm 2.77 ) \times 10^{-4}$ \\ 
$h_0^{(2)}$ & $ ( -3.97 \pm 4.45 ) \times 10^{-5}$    &  $ (8.55 \pm 4.12 ) \times 10^{-5}$ \\ 
\hline
\end{tabular} 
}
\caption{Hadronic power corrections  fit to BR$(B\to K^* \gamma)$, ${\rm BR}(B^+ \to K^{+*} \mu^+ \mu^-)_{q^2 \in [1.1-6.0]\;{\rm GeV}^2}$ and the $B \to K^* \mu^+ \mu^-$ 
observables in the low $q^2$ bins up to 8 GeV$^2$.
\label{tab:imp1_set78_Had}
}
\end{center} 
\end{table}

We used the same set of observables to make one and two operator NP fits to $\delta C_{9}$ and $\delta C_{7,9}$  
assuming the Wilson coefficients to be either real or complex in Table~\ref{tab:C79lowq2}.
Interestingly the real parts of the best fit point for $\delta C_9$ in all four cases are compatible within their 68\% confidence level and all these NP scenarios
have a  better description of the data compared to the SM hypothesis with larger than $4\sigma$ significance.
The fits suggest sizeable imaginary parts for the Wilson coefficients, with rather large uncertainties.
In principle considering the CP asymmetric observables of the $B \to K^* \ell^+ \ell^-$ decay should allow us to further constraint the 
imaginary parts of the Wilson coefficients but the current experimental data on the relevant CP-asymmetric observables~\cite{Bobeth:2008ij,Altmannshofer:2008dz} 
such as $A_{7,8,9}(B\to K^* \mu^+ \mu^-)$\footnote{For the correct sign of $J_{7,8,9}$ see Ref.~\cite{Gratrex:2015hna} where the information on the helicity angle $\phi$ is unambiguous.} are not stringent enough to put any significant constraints on the imaginary parts.

\begin{table}[t!]
\ra{1.}
\rb{1.3mm}
\begin{center}
\setlength\extrarowheight{2pt}
\begin{tabular}{|c|c|c|}
\hline
  \multicolumn{2}{|c}{up to $q^2=8$ GeV$^2$ obs.} & \multicolumn{1}{c|}{($\chi^2_{\rm SM}=54.9$)  }       \\  
\hline
 & \multicolumn{1}{c}{best fit value} & \multicolumn{1}{|c|}{$\chi^2_{\rm min}$}\\
 \hline \hline
$\delta C_9$ & $ -1.15 \pm 0.22 $ & $38.1$  \\  
\hline \hline
$\delta C_7$ & $  0.04 \pm 0.03 $ & \\[-6pt]
\footnotesize{\&} & & $36.0$ \\  [-4pt]
$\delta C_9$ & $ -1.47 \pm 0.31 $ &  \\
\hline
\end{tabular}    \quad \quad \quad 
\begin{tabular}{|c|c|c|}
\hline
  \multicolumn{2}{|c}{up to $q^2=8$ GeV$^2$ obs.} & \multicolumn{1}{c|}{($\chi^2_{\rm SM}=54.9$)  }       \\  
\hline
 & \multicolumn{1}{c}{best fit value} & \multicolumn{1}{|c|}{$\chi^2_{\rm min}$}\\
 \hline \hline
$\delta C_9$ & $(-1.03 \pm 0.25) + i(-2.04 \pm 0.58)  $ &  $33.9$  \\                                                         
\hline  \hline  \rule{0pt}{3ex} 
$\delta C_7$ & $(  0.03 \pm 0.03) +  i( 0.09 \pm 0.05) $ & \\  [-6pt]
\footnotesize{\&} & & $30.3$ \\  [-4pt]
$\delta C_9$ & $( -1.30 \pm 0.35) +  i(-2.40 \pm 0.73) $ & \\ 
\hline
\end{tabular}    
\caption{One and two operator NP fits for real (complex) $\delta C_9$ and $\delta C_{7,9}$ 
on the left (right) considering the same observables as mentioned in the caption of Table~\ref{tab:imp1_set78_Had}.
\label{tab:C79lowq2}
}
\end{center} 
\end{table}

As shown in appendix~\ref{sec:q2behaviour}, the effect of NP  contributions to  observables from $C_7$ and $C_9$ can be embedded in the more general case of the hadronic contributions. 
Due to the embedding, any lepton flavour universal NP  contribution to the Wilson coefficients, $C_7$ and $C_9$ can be simulated by some hadronic effect and it is not possible to rule out underestimated hadronic explanation in favour of the NP one by only considering CP-averaged $B\to K^* \mu^+ \mu^-$ observables\footnote{In principle, the embedding can be broken even with flavour universal NP  contribution to the Wilson coefficients, $C_7$ and $C_9$ since the imaginary parts in Wilson coefficients correspond to CP-violating ``weak'' phases while the imaginary parts of the hadronic contributions correspond to CP-conserving ``strong'' phases~\cite{Bobeth:2008ij,Altmannshofer:2008dz,Bobeth:2011gi,Bobeth:2011nj}. However, the current data on CP-asymmetric observables for $B\to K^* \mu^+ \mu^-$~\cite{Aaij:2014bsa,Aaij:2015oid} and $B\to K^* \gamma$~\cite{Coan:1999kh,Nakao:2004th,Aubert:2009ak,Aaij:2012ita,Amhis:2014hma} are not constraining enough to allow us to make this differentiation.}. 
However, there can be a statistical comparison between the NP fit versus the hadronic contribution fit 
since the embedding of $C_{7,9}^{\rm NP}$ contributions in the hadronic ones results in nested scenarios which can be statistically compared.
Employing the Wilks' theorem, two nested scenarios can be compared by considering the difference in the minimum $\chi^2$ of each scenario and the difference between the number of parameters for each scenario.
In Table~\ref{tab:WilksTest_8GeV2} the significance of the improvement of the fit 
in the hypothesis with more free parameters has been compared to the ones  with less free parameters using the Wilks' theorem.
Clearly, the scenario of real valued NP contribution for each Wilson coefficient is nested in the NP scenario with the same Wilson coefficient with complex values.
Moreover, as shown in appendix~\ref{sec:q2behaviour}, NP contributions in $C_7$ and/or $C_{9}$ can be considered as a nested scenario with respect to non-factorisable hadronic contributions. In the last column in Table~\ref{tab:WilksTest_8GeV2} which describes the improvement of the hadronic fit with respect to the NP fit, we have considered the case where all 18 parameters describing the hadronic fit (Eq.~(\ref{eq:hlambda}) and Eq.~(\ref{eq:hzero})) are free parameters and the result has been compared to the NP fits which are equivalent to only having $1,2$ or 4 parameters in $h_{+,-,0}^{(0,1,2)}$ to be free.
While the hadronic solution and the NP explanation both have a better description of the measured data with a significance of larger than $3\sigma$, there is always less than $1.5\sigma$ improvement when going from the NP fits to the hadronic one.
Compared to the scenarios with real contributions to $C_9$ or $C_{7,9}$, the NP fit has $\sim 2\sigma$ improvement when the Wilson coefficients are considered to be complex. 

\begin{table}[h!]
\ra{1.2}
\rb{1.3mm}
\begin{center}
\setlength\extrarowheight{2pt}
\scalebox{0.9}{
\begin{tabular}{|l|c|c|c|c|c|c|c|}
\hline
nr. of free parameters		    & 1			  & 2 				   & 2 		            & 4 			        & 18 \\ [-6pt]
				    & (Real $\delta C_9$) & (Real $\delta C_7,\delta C_9$) & (Complex $\delta C_9$) & (Complex $\delta C_7,\delta C_9$) & (Complex $h_{+,-,0}^{(0,1,2)}$)\\
\hline
0 (plain SM)                        & $4.1\sigma$      & $4.0\sigma$ & $4.2\sigma$ 	& $4.1\sigma$ 	& $3.1\sigma$\\
1 (Real $\delta C_9$)               & --               & $1.5\sigma$ & $2.1\sigma$	& $2.0\sigma$	& $1.5\sigma$\\
2 (Real $\delta C_7,\delta C_9$)    & --               & --          & --		& $1.9\sigma$	& $1.4\sigma$\\
2 (Complex $\delta C_9$)            & --               & --          & -- 		& $1.4\sigma$	& $1.1\sigma$\\
4 (Complex $\delta C_7,\delta C_9$) & --               & --          & --		& --		& $0.95\sigma$\\
\hline
\end{tabular} 
}
\caption{Improvement of the hadronic fit and the scenarios with real and complex NP  contributions to Wilson coefficients  $C_7$ and $C_9$ compared to the SM hypothesis and compared to each other.
\label{tab:WilksTest_8GeV2}
}
\end{center} 
\end{table}  

The slight differences of Table~\ref{tab:imp1_set78_Had}-\ref{tab:WilksTest_8GeV2} compared to the relevant similar results of Ref.~\cite{Chobanova:2017ghn} are due to the modified parameterisation of the hadronic contributions for $\lambda=0$, and also due to the inclusion of two additional observables, BR($B\to K^* \gamma$) and BR$(B^+ \to K^{+*} \mu^+ \mu^-)_{q^2 \in [1.1-6.0]\;{\rm GeV}^2}$.
Nonetheless, the conclusion remains the same; adding 14-17 more parameters compared to the NP  fit does not significantly improve the fit (although the improvement of the hadronic fit compared to NP one is now slightly larger).

The results indicate preference for rather large imaginary parts in the fit parameters which is the consequence of not having included CP asymmetric observables in our fits as the available experimental results on such observables are not very constraining at present.
Thus, at the moment the statistical comparison favours the NP explanation and more constraining data on CP-asymmetric observables would be needed to determine whether it should be real or complex. However, the situation remains inconclusive. With the set of observables considered in this analysis, the NP fit can be embedded in the hadronic fit. In this sense one cannot disprove the hadronic option in favour of the NP one as discussed above.

With the present results, there is no indication that higher powers of $q^2$ than the ones which are attainable by NP contributions to  $C_7$ and $C_9$ would be required to explain the $B \to K^* \mu^+ \mu^-$ data. 
However, this might be due to the size of the current $q^2$ bins which can potentially smear out a significant $q^2$ dependence and thus smaller binning can shed more light on this issue\footnote{The LHCb has provided a finer binning using the method of moments where the bins have a range of $\sim\!1$ GeV$^2$, but compared to the results with larger bins obtained by the maximum likelihood method, the experimental uncertainties are currently much larger.}.
Moreover, an unbinned analysis may show a hadronic
structure which is hidden in the present data due to the large bins and the release of unbinned data to the theory community could potentially clear up this issue (see also Refs.~\cite{Bobeth:2017vxj,Blake:2017fyh,Hurth:2017sqw,Mauri:2018vbg}).

In principle, one could have higher powers of $q^2$ in the parameterisation of the hadonic contributions of Eqs.~(\ref{eq:hlambda})~and~(\ref{eq:hzero}) and it would clearly still keep the embedding of NP contributions. 
In fact, if a fit would show preference for such higher power terms (e.g. $h_\lambda^{(3)}$) it would indicate that the $B \to K^* \mu^+ \mu^-$ data would be best described by underestimated power corrections since NP contributions would not be able to mimic such $q^2$ terms. 
However, as can be seen from Table~\ref{tab:imp1_set78_Had}, the fitted parameters $h_\lambda^{(0,1,2)}$ are almost all compatible with zero, within the $1\sigma$ range, which is due to the fact that the current data on the $B \to K^* \mu^+ \mu^-$ decay are not constraining enough to further constrain the 18 free parameters.
Hence including $h_\lambda^{(3)}$ terms in the parameterisation would results in 24 unknown parameters and consequently even looser constraints on the fitted parameters.
Furthermore, since the Wilks' test indicates that  helicity- and $q^2$-dependent terms beyond the NP contribution from $C_9$ are not statistically preferred, it can be understood that adding higher powers of $q^2$ in the power correction ansatz would not change our conclusion.

\section{Theoretical estimations of the hadronic contributions}\label{sec:implementations}
The short-distance NP contributions due to $\delta C_9^{\rm NP}$ (and/or $\delta C_7^{\rm NP}$) 
can be mimicked by long-distance effects in $h_\lambda$.
Therefore, a proper estimation of the size of the hadronic contributions is highly desirable and crucial in determining whether the observed anomalies in $B\to K^* \mu^+ \mu^-$ observables
result in a significant NP interpretation. There are different approaches offered in the literature in order to estimate the hadronic contributions, on which we elaborate below.

\subsection{Various approaches}
In the ``standard'' method the hadronic contributions are estimated using the QCD factorisation  formalism where the factorisable as well as non-factorisable contributions from vertex corrections~\cite{Asatrian:2001de,Asatryan:2001zw}, weak annihilation and spectator scattering~\cite{Beneke:2001at,Beneke:2004dp} are taken into account.
However,  higher powers of ${\cal O}(1/m_b)$ remain unknown within the QCDf formalism. 
In the so-called ``full form factor'' method (see i.e. Ref.~\cite{Hurth:2016fbr}), only the power corrections to the non-factorisable piece in the QCDf formula are not known and are usually guesstimated to be $10\%$, $20\%$ or even higher percentages compared to the leading non-factorisable  contributions.
For the $B \to K^* \ell^+ \ell^- $ observables, as well as other exclusive  $B \to V(P) \ell^+ \ell^-$ decays with a vector (pseudoscalar) meson in the final state which appear in the global fit of section~\ref{sec:allWilson} we have used this ``standard'' method with a 10\% assumption for the power corrections.

Among the hadronic contributions, the most relevant ones are due to the charm loops arising from the current-current operators $O_{1,2}$.
The power corrections relevant to these charm loops, the soft gluon effects, have been estimated in Ref.~\cite{Khodjamirian:2010vf} using the LCSR formalism in the $q^2\lesssim 1\text{ GeV}^2$ region where $q^2 \ll 4 m_c^2$ holds.
The results are  extrapolated up to the $J/\psi$ resonance by employing dispersion relations and using the experimental data from $B \to J/\psi K^*$ and $B \to \psi(2S) K^*$ decays.
However, in the theoretical input of the dispersion relation the leading order non-factorisable effects (available from QCDf calculations~\cite{Beneke:2001at,Beneke:2004dp}) which have an important contribution to the analyticity structure are not included. 
Moreover, the phases of the resonant amplitude relative to the short-distance contribution for each of the three amplitude structures (for both resonances) are just set to zero.

It is claimed in Ref.~\cite{Khodjamirian:2012rm} that for the $B\to K^* \ell^+ \ell^-$ decay  hadronic contributions from the $s$ quark (i.e. the $\phi$ meson pole) have a $1/q^2$ factor for the transverse polarisation and hence get enhanced at small $q^2$. 
Therefore, in order to have a precise estimation it would be preferable to have separate dispersion relations for the resonances due to the $c$ quark and due to the $s$ and $b$ quarks. This has not been done since the theory calculations for some of the relevant contributions (e.g. Refs.~\cite{Asatrian:2001de,Asatryan:2001zw}) are not available in a flavour separated way.

One way to compensate the missing leading order factorisable corrections in the Khodjamirian et al. method is to just add these missing contributions to the phenomenological model.
However, the theoretical error which enters this procedure is unclear. This is done for example in Ref.~\cite{Ciuchini:2017mik}, while the subleading hadronic contributions have been accounted for 
by considering the phenomenological description of Ref.~\cite{Khodjamirian:2010vf} valid up to $q^2\lesssim 9\text{ GeV}^2$ (referred to as PMD in Ref.~\cite{Ciuchini:2017mik}). 

In Ref.~\cite{Bobeth:2017vxj}, the most promising approach to the hadronic contributions is offered, which may lead to a clear separation  of hadronic and NP effects. The authors consider the analyticity of the amplitude.
Building upon the work of Refs.~\cite{Khodjamirian:2010vf,Khodjamirian:2012rm}, 
both the leading and subleading hadronic contributions arising from the  charm loop 
contributions of the current-current operators $O_{1,2}$ have been estimated.
The calculations are performed at $q^2<0$ where the theory predictions 
for the leading terms in QCDf~\cite{Asatrian:2001de,Asatryan:2001zw,Beneke:2001at,Beneke:2004dp} 
as well as the subleading terms in LCSR~\cite{Khodjamirian:2010vf,Khodjamirian:2012rm} are reliable  
and in combination with the experimental information on the  $B \to J/\psi K^*$ and $B \to \psi(2S) K^*$ decays,
the hadronic contributions due to the charm loops are estimated in the physical region up to the $\psi(2S)$ resonance. They use the well-known $z$ parameterisation (see e.g.~\cite{Boyd:1995cf,Bourrely:2008za,Na:2010uf})\footnote{Most recently (in Ref.~\cite{Chrzaszcz:2018yza}), the authors have analysed the convergence of the $z$ expansion in great detail. We still use their 
explicit results based on the expansion up to the $z^2$ terms.}.

The authors of this paper argue that the cut giving rise to light hadron resonances can be neglected due to suppression by the nonperturbative Okubo-Zweig-Iizuka (OZI)
rule~\cite{Okubo:1963fa,Zweig:1964jf,Iizuka:1966fk} both above and below the $\phi$ resonance as long as the effects of the $\phi$ are not resolved, e.g. if an appropriate binning is applied~\cite{vanDykPrivateComm}.
One may conclude from this argument that the separation of the dispersion relation for $c$ and $b/s$  contributions as proposed in Ref.~\cite{Khodjamirian:2012rm} (see above) is phenomenologically not necessary. 

Unfortunately, the correlations among the theoretical uncertainties of the complex parameters describing the parameterisation of the hadronic contributions have not been provided in Ref.~\cite{Bobeth:2017vxj}.
Nonetheless, the uncertainties of each of the parameters are available which used without the correlations leads to a very conservative theory estimation of the hadronic contributions.

Finally, in Ref.~\cite{Blake:2017fyh}, all the hadronic contributions, from the charm (and light quark) resonances are modeled as Breit-Wigner amplitudes.
The effect of the $J/\psi$ and $\psi(2S)$ (and the rest of the) resonances on $B \to K^* \mu^+ \mu^-$ 
observables is estimated (up to an overall global phase for each resonance) using measurements on the branching fractions and polarisation amplitudes of the resonances.
The overall phase can be assessed from simultaneous fits to the short- and long-distance components in the $K^* \mu^+ \mu^-$ final states. 
However, since  this measurement is currently not available, in Ref.~\cite{Blake:2017fyh} 
all possible values for the overall phase of each resonant state have been assumed and therefore the results are rather unconstraining.
The theory predictions of both Ref.~\cite{Khodjamirian:2010vf} and Ref.\cite{Bobeth:2017vxj} can be reproduced with appropriate choices for the unknown parameters entering the empirical model.

\subsection{Comparison of the different approaches}
\begin{figure}[h!]
\centering
\includegraphics[width=0.48\textwidth]{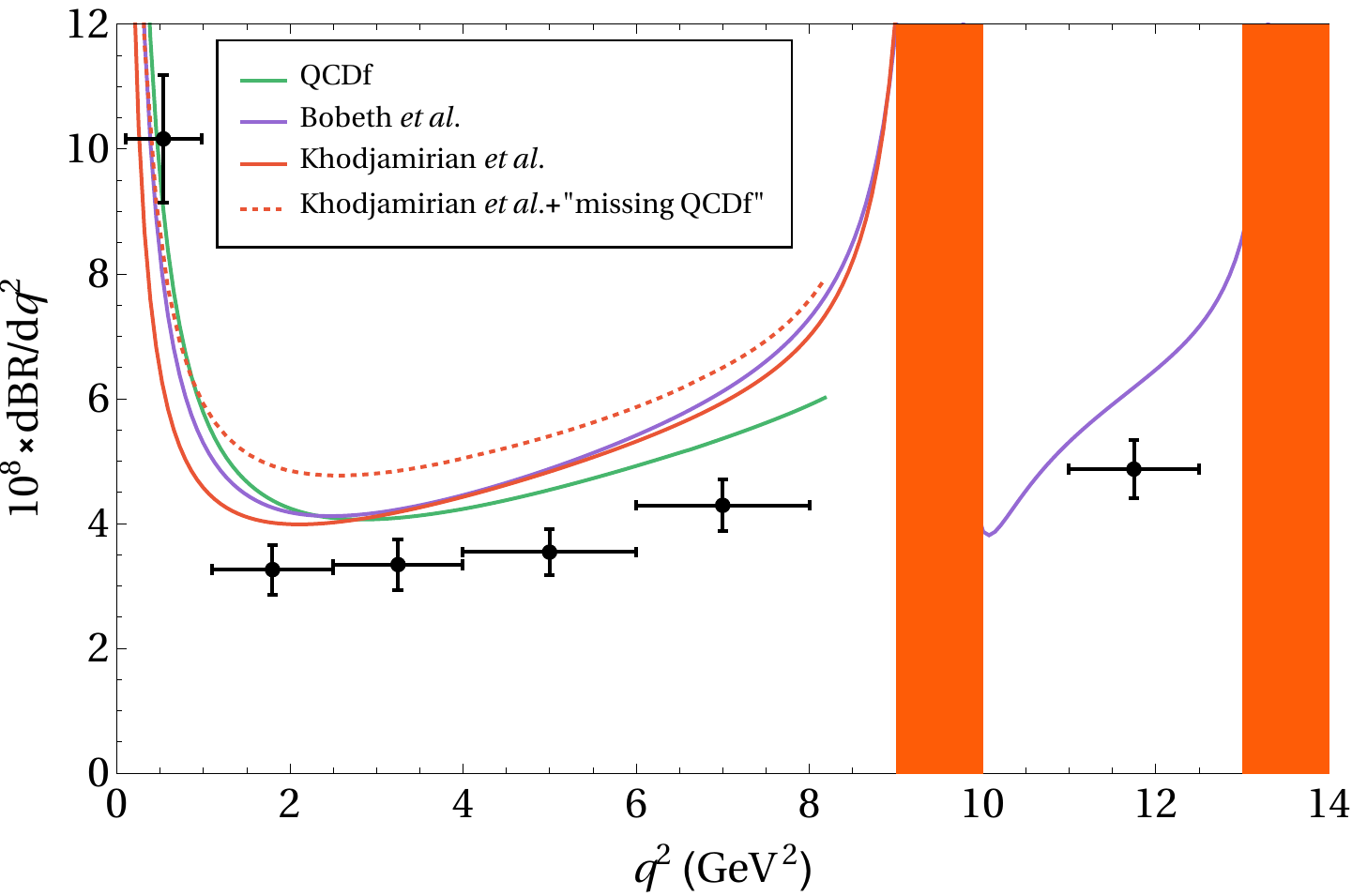}
\includegraphics[width=0.48\textwidth]{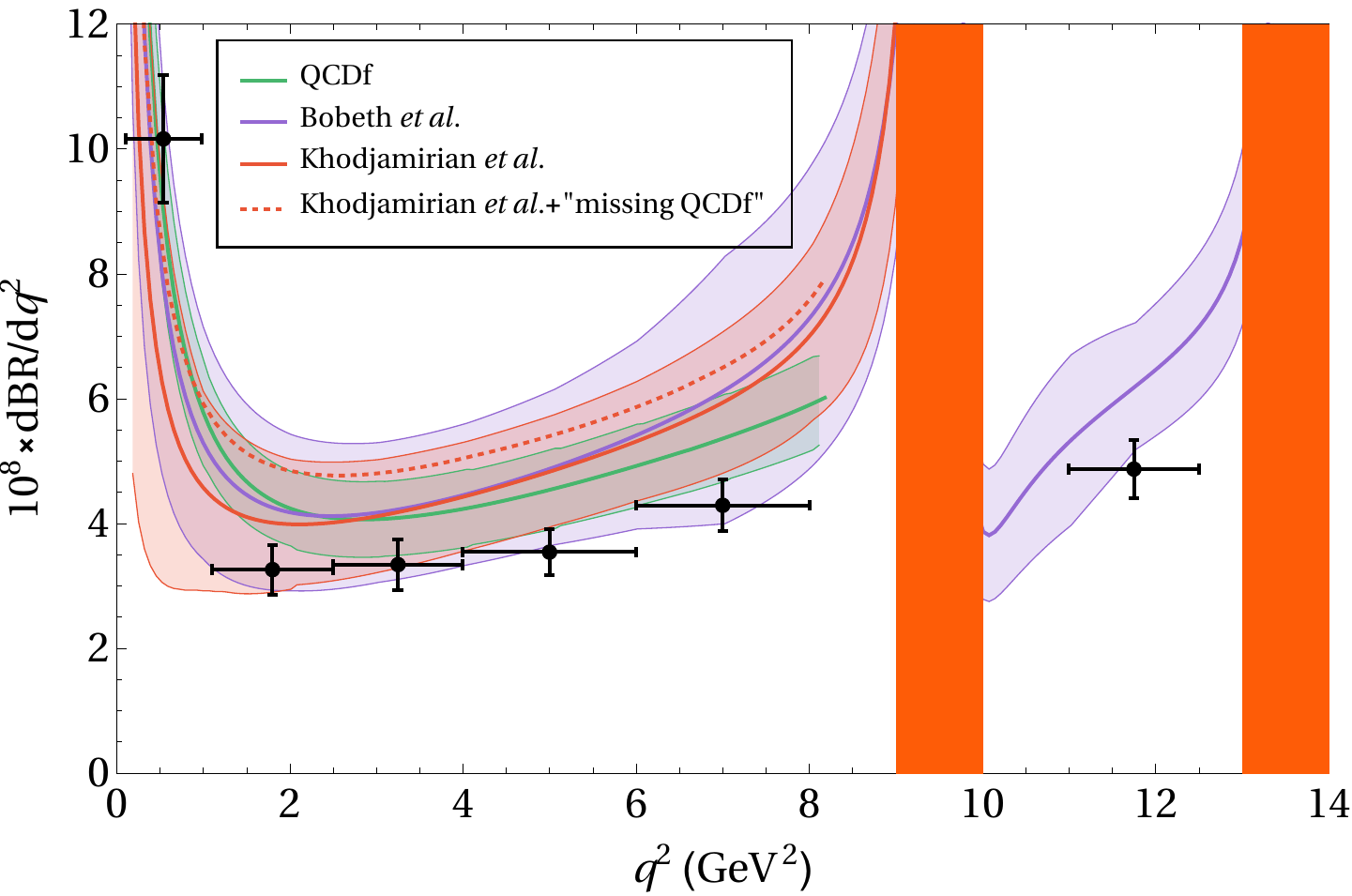}
\caption{The SM predictions of $d{\rm BR}(B\to K^* \mu^+ \mu^-)/dq^2$  within various 
implementations of the hadronic contributions without (with) the theory uncertainties on the left (right). 
For the ``QCDf'' implementation the full form factor method  has been considered,
with a 10\% error assumption for the power corrections.
The theory error of the Khodjamirian et al. implementation is obtained by considering the relevant parameter uncertainties that goes into the phenomenological formula.
For the theoretical uncertainty of the Bobeth et al. method the correlations of the parameters describing the hadronic contributions have not been used.
The theoretical uncertainty of the method where the leading order non-factorisable contributions are added to the phenomenological model of Ref.~\cite{Khodjamirian:2010vf} (Khodjamirian et al. + ``missing QCDf'') are not shown.\label{Fig:BR_implementations}}
\end{figure}

\begin{figure}[h!]
\centering
\includegraphics[width=0.48\textwidth]{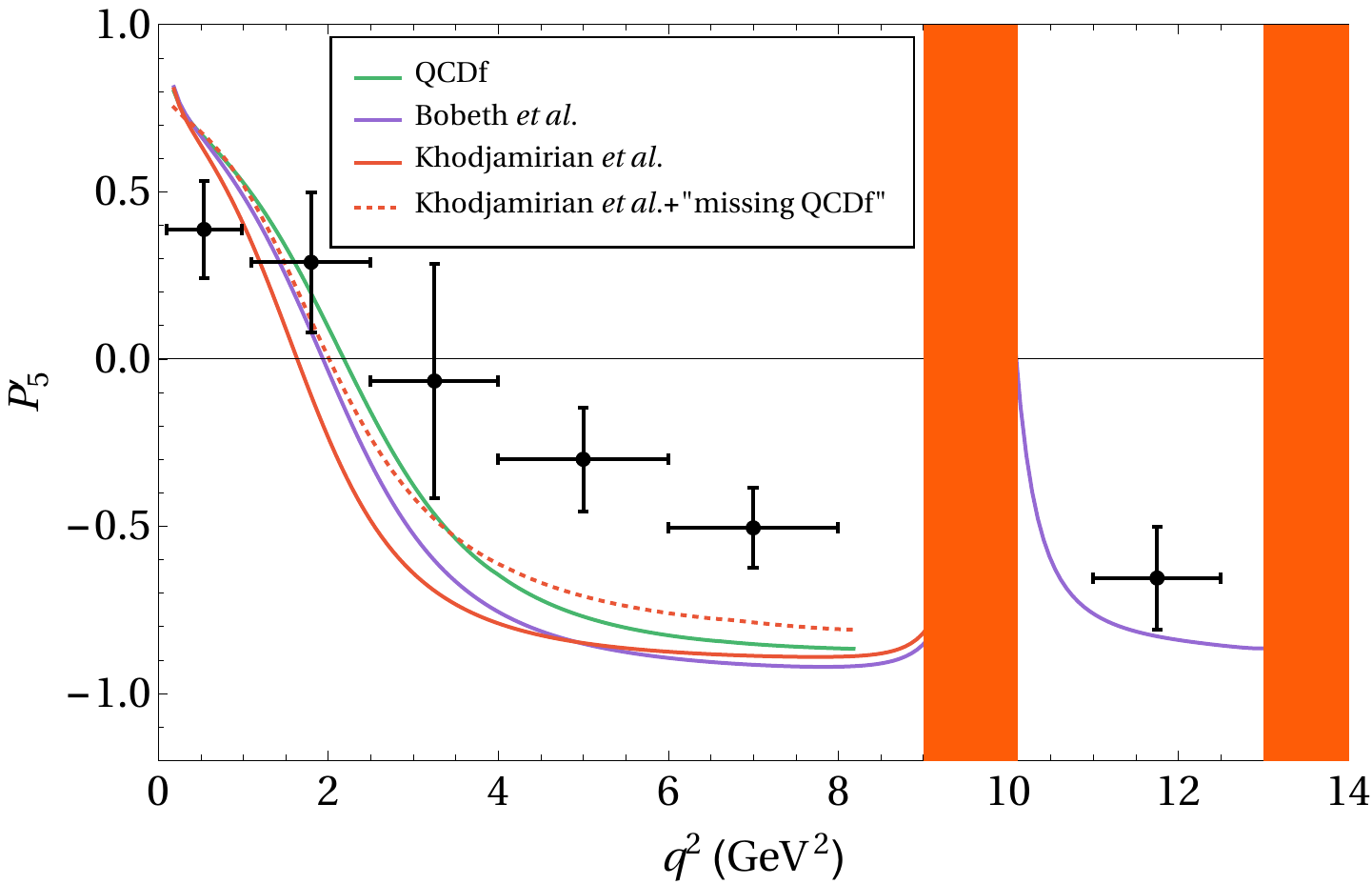}
\includegraphics[width=0.48\textwidth]{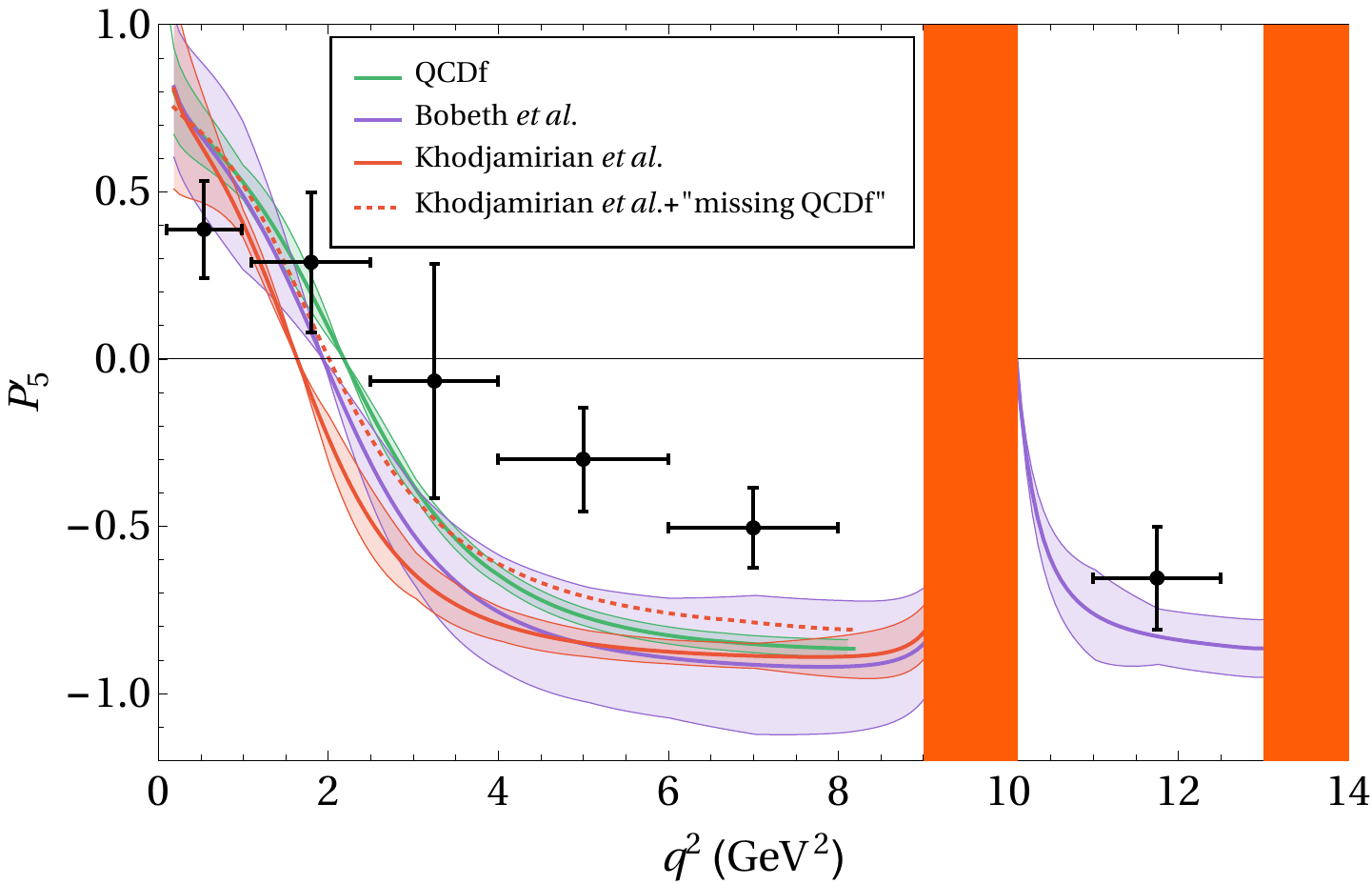}
\caption{The SM predictions of $P_5^\prime(B\to K^* \mu^+ \mu^-)$ within various 
implementations of the hadronic contributions as described in the caption of Fig.~\ref{Fig:BR_implementations}.\label{Fig:5p_implementations}}
\end{figure}

To show how the various theory estimations differ in their predictions of $B\to K^* \mu^+ \mu^-$ observables, 
the SM results for $d{\rm BR}/dq^2$ and $P_5^\prime$  using the various implementations of the hadronic contributions are given in Fig.~\ref{Fig:BR_implementations} and Fig.~\ref{Fig:5p_implementations}, respectively.
In the ``standard'' method, the predictions are given for below $q^2 = 8$ GeV$^2$ where QCDf calculations are reliable while the phenomenological model of Khodjamirian et al.  
is considered up to $q^2 < 9$ GeV$^2$ and only the Bobeth et al. method has a prediction for also between the $J/\psi$ and $\psi(2S)$ resonances.
Interestingly the central values of the latter two methods increase the tension with experimental measurement for both $d{\rm BR}/dq^2$ and $P_5^\prime$ and it seems that the contribution from the power corrections tends to further escalate the tension with the data. The theory errors of these predictions, however, are larger (for the Bobeth et al. method this is due to the lack of correlations among uncertainties, which are not given in Ref.~\cite{Bobeth:2017vxj}).
From Figs.~\ref{Fig:BR_implementations} and \ref{Fig:5p_implementations} it can be seen that the $B\to K^* \mu^+ \mu^-$ observables (BR and $P_5^\prime$) within the various available methods for estimation of the non-factorisable corrections are in agreement at the $1\sigma$ level.


The significance of the NP interpretation for the $B\to K^* \mu^+ \mu^-$ anomalies clearly depends on the theory estimations of the hadronic contributions.
In Table~\ref{tab:NP_implementations} the significance of different NP scenarios (for one operator fits to $\delta C_{7}$, $\delta C_{9}$ or $\delta C_{10}$) are given using the ``standard'' implementation (with 10\% error assumption on the power corrections)  and the Bobeth et al. implementation of the non-factorisable corrections.
While in both implementations the NP  contribution to $C_9$ constitutes the favoured scenario, the significance and the best fit values are different. 
Nevertheless, one finds consistency at the $2\sigma$ level~\cite{Hurth:2017hxg} of the deviations in the angular observables and the deviations found in the measurements in the ratios $R_K$ and $R_{K^*}$, which is a tantalising hint for NP.

\begin{table}[t!]
\ra{1.2}
\rb{1.3mm}
\begin{center}
\setlength\extrarowheight{2pt}
\scalebox{0.9}{
\begin{tabular}{|l|c||c|c||c|c||c|c|}
\hline
                    & SM             & \multicolumn{2}{c||}{$\delta C_7$} & \multicolumn{2}{c||}{$\delta C_9$} & \multicolumn{2}{c|}{$\delta C_{10}$} \\ \cline{2-8}
                    & b.f. value     & b.f. value & $\chi^2_{\rm min}$    & b.f. value & $\chi^2_{\rm min}$    & b.f. value & $\chi^2_{\rm min}$    \\
\hline
QCDf                & 60.9          & $-0.03 \pm 0.02$ & $58.9 (1.4\sigma)$ & $-1.05 \pm 0.21$ & $45.4 (3.9\sigma)$ & $-0.17 \pm 0.35$ & $60.7 (0.5\sigma)$  \\
Bobeth et al.       & 54.8          & $-0.03 \pm 0.03$ & $53.5 (1.1\sigma)$ & $-1.26 \pm 0.28$ & $43.9 (3.3\sigma)$ & $0.48 \pm 0.63$ & $54.1 (0.8\sigma)$  \\
\hline
\end{tabular} 
}
\caption{The $\chi^2$ of the one operator NP fit compared to the SM within the ``standard'' QCDf method (with a 10\% error assumption on the power corrections) and the Bobeth et al. method.
The observables considered in the fit include BR($B \to K^* \gamma$), BR($B^+\to K^{+*} \mu^+ \mu^-$) in the [1.1-6.0] and [15-19] GeV$^2$ bins and all the $B\to K^* \mu^+ \mu^-$ observables in both  high and low $q^2$ bins. 
\label{tab:NP_implementations}
}
\end{center} 
\end{table}  

\section{Fit to NP including scalar \& pseudoscalar operators}\label{sec:allWilson}

\begin{figure}[h!]
\centering
\includegraphics[width=0.48\textwidth]{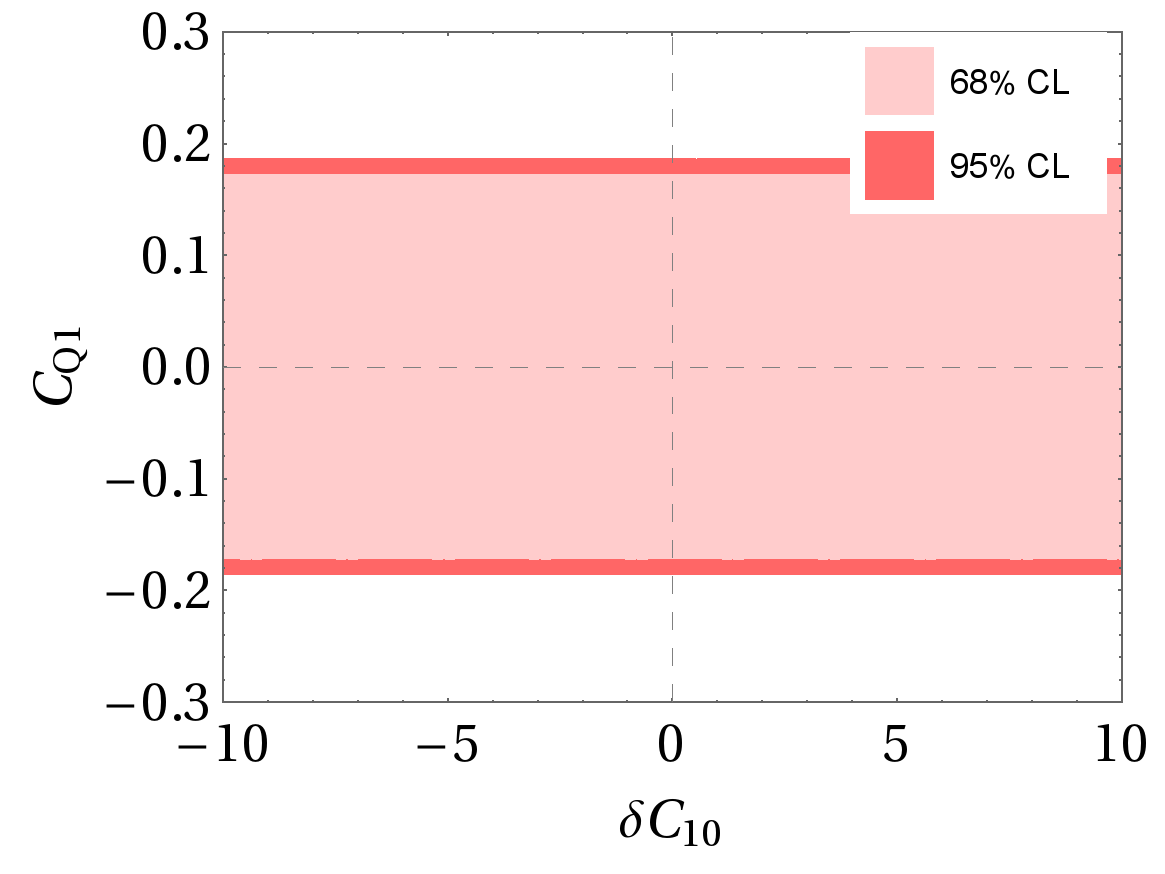}
\includegraphics[width=0.48\textwidth]{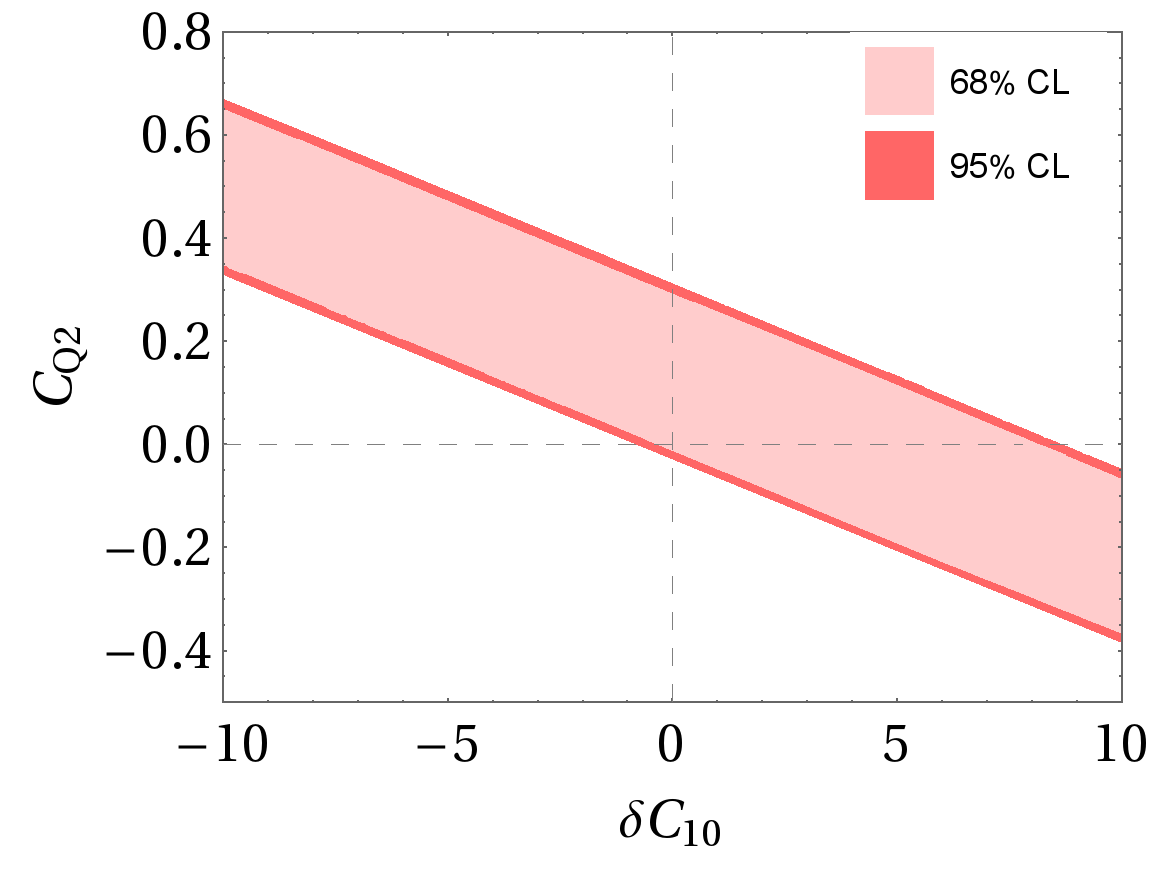}
\caption{Two-dimensional projection of the three operator fit to $C_{10}$, $C_{Q_{1}}$ and $C_{Q_{2}}$. The (light) red contours correspond to the (68) 95\% C.L. regions.
\label{Fig:C10CQ1CQ2_3op}}
\end{figure}

Assuming the observed tensions in $b \to s \ell^+ \ell^-$ data to be due to NP  contributions there is in principle no reason why NP contributions should affect only one or two Wilson coefficients. In particular, a complete NP scenario incorporates many new particles and can have extended Higgs sector, affecting the Wilson coefficients $C_{7\cdots10}$ and requiring scalar and pseudoscalar contributions. It is often considered that the data on BR($B_s \to \mu^+ \mu^-$) remove the possibility to have large scalar and pseudoscalar Wilson coefficients $C_{Q_{1,2}}$ (see e.g. Ref.~\cite{Mahmoudi:2012un} for the definition of the relevant operators). While this is rather true for $C_{Q_1}$, there exists a degeneracy between $C_{10}$ and $C_{Q_2}$ which makes it possible to have simultaneously large values for both Wilson coefficients. To demonstrate this, we perform a fit to BR($B_s \to \mu^+ \mu^-$) when the three Wilson coefficients $C_{10,{Q_{1,2}}}$ are varied independently. The results can be seen in Fig.~\ref{Fig:C10CQ1CQ2_3op}, where two dimensional projections of the constraints by BR($B_s \to \mu^+ \mu^-$) are shown on the $C_{10,{Q_{1,2}}}$ Wilson coefficients. While $C_{Q_{1}}$ is still limited between $\pm 0.2$, both $C_{10}$ and $C_{Q_{2}}$ can have large values, due to the compensation in the BR($B_s \to \mu^+ \mu^-$) formula (see e.g. Ref.~\cite{Mahmoudi:2012un}). 

\begin{figure}[h!]
\centering
\includegraphics[width=0.48\textwidth]{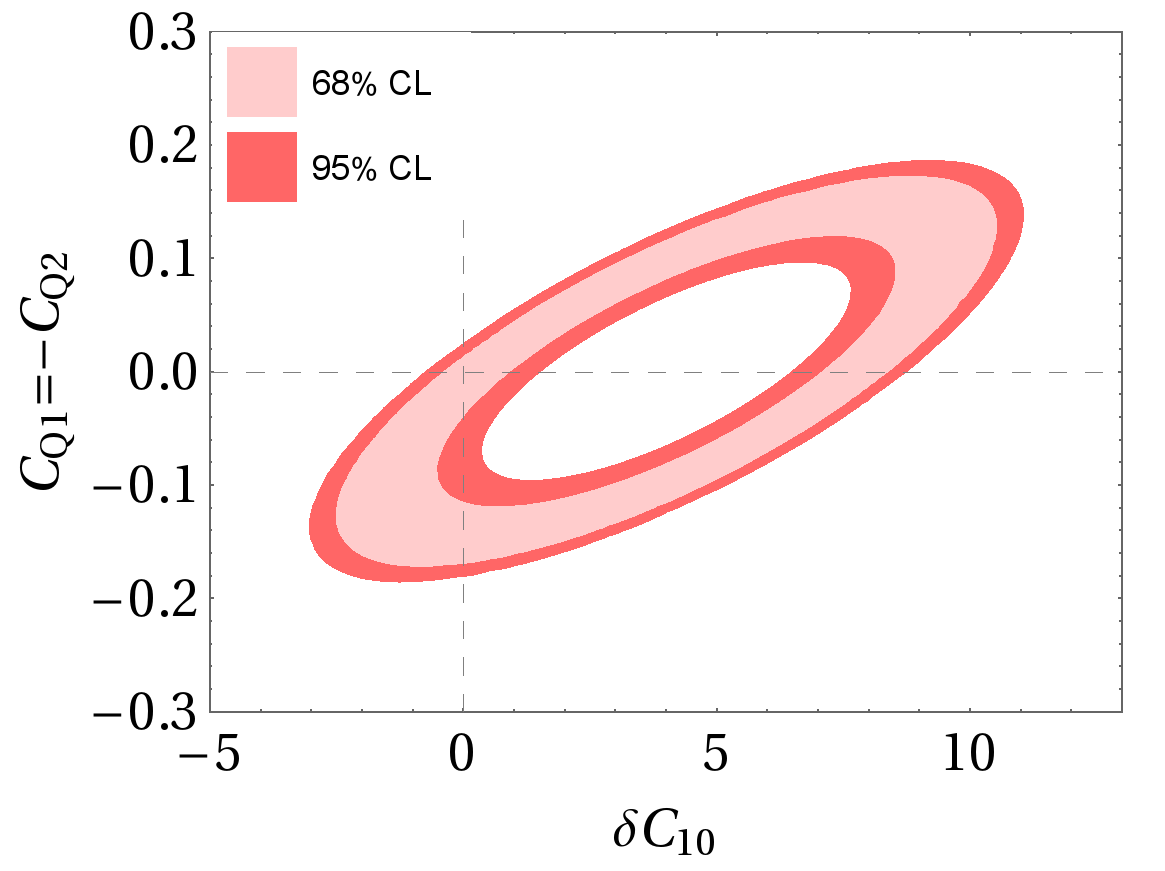}
\caption{Two operator fit to the BR($B_s \to \mu^+ \mu^-$) with NP  contributions in  $C_{10}$ and $C_{Q_{1}}=-C_{Q_{2}}$.
The (light) red contours correspond to the (68) 95\% C.L. regions. 
\label{Fig:C10CQ_2op}}
\end{figure}

We also consider  the case where $C_Q \equiv C_{Q_1}=-C_{Q_2}$\footnote{This relation arises for example in the SMEFT framework assuming a SM Higgs.}, and $C_{10}$ are varied separately. The results are shown in Fig.~\ref{Fig:C10CQ_2op}. In such a case the degeneracy between $C_{Q_2}$ and $C_{10}$ is broken, and the scalar and pseudoscalar contributions are limited between $\pm 0.2$. It is remarkable that $\delta C_{10}$ can take large values, whereas it is limited between $\pm 1$ (or $[+7, +9]$) resulting in $3\lesssim|C_{10}^{\rm SM+NP}|\lesssim5$ when $C_{Q_1}$ and $C_{Q_2}$ are set to zero.

As a consequence, the branching ratio of $B_s \to \mu^+ \mu^-$ cannot be used to set simultaneously strong constraints on $C_{10}$ and $C_{Q_{1,2}}$ in generic NP scenarios, but can only be used to justify why $C_Q$ or $C_{Q_1}$ can have very limited contributions. 
Conversely, while the measurement of the branching ratio of $B_s \to \mu^+ \mu^-$ is used to justify why the scalar and pseudoscalar contributions are set to zero in a specific fit, 
it \emph{cannot} be used to set constraints on $C_{10}$ any more since there is a cancellation between $C_{10}$ and $C_{Q_2}$ and to rule out scalar and pseudoscalar contributions 
by only considering the branching ratio of $B_s \to \mu^+ \mu^-$ requires the assumption that $\delta C_{10}$ is small which is only justified when doing a global fit to all $b \to s \ell^+ \ell^-$ data (see Ref.~\cite{Beaujean:2015gba} for a detailed study of how the scalar and tensorial Wilson coefficients are constrained when considering all relevant  $b \to s \ell^+ \ell^-$ data).

We have thus expanded our study to include NP in the global fit\footnote{The full list of observables considered in this study can be found in Ref.~\cite{Hurth:2016fbr} where for ${d{\rm BR}/dq^2}(B\to K^{*} \mu^+ \mu^-)$ the binning of Ref.~\cite{Aaij:2016flj} with 3 fb$^{-1}$ data has been considered
and  BR($B\to K^* \gamma$), $R_{K^*}$ and the angular observable $F_H$ of the $B^+ \to K^+ \mu^+ \mu^-$ decay~\cite{Aaij:2014tfa} have also been added to the global fit.} 
to $b \to s \ell^+ \ell^-$ from other Wilson coefficients besides $C_7$ and $C_9$ to also include the chromomagnetic operator as well as the axial-vector, scalar and pseudoscalar operators with overall 10 independent Wilson coefficients $C_7, C_8, C_9^{\ell}, C_{10}^{\ell}, C_{Q_1}^\ell, C_{Q_2}^\ell$
(assuming lepton flavours to be $\ell=e,\mu$). 
Considering the operators where the chirality of the quark currents are flipped (primed Wilson coefficients), there will be 20 free parameters\footnote{Since the experimental data on CP-asymmetric observables cannot put stringent constraints on the imaginary parts of the complex Wilson coefficients, we have only considered real values for the Wilson coefficients.}.

To perform our fits, the theoretical correlations and errors are computed using {\tt SuperIso v4.0}, which incorporates an automatic multiprocessing calculation of the covariance matrix for each parameter point.  We have considered a 10\% error assumption for the power corrections. The experimental correlations are also taken into account. The Minuit library \cite{James:1975dr} has been used to search for the global minima in high dimensional parameter spaces. For each fit we carefully searched for the local minima in order to find the global minima.

\begin{table}[th!]
\begin{center}
\setlength\extrarowheight{3pt}
\hspace*{-1.cm}
\scalebox{0.90}{
\begin{tabular}{|l|r|r|c|}
\hline 
 \multicolumn{4}{|c|}{All observables  ($\chi^2_{\rm SM}=118.8$)} \\ \hline
                          & b.f. value & $\chi^2_{\rm min}$ & ${\rm Pull}_{\rm SM}$  \\ 
\hline \hline
$\delta C_{7} $          	& $ 	-0.01	\pm	0.01	 $ & $ 	117.9	 $ & $	0.9	\sigma	 $  \\
$\delta C_{Q_{1}}^{\mu} $  	& $ 	-0.03	\pm	0.08	 $ & $ 	118.7	 $ & $	0.3	\sigma	 $  \\
$\delta C_{Q_{1}}^{e} $    	& $ 	-0.91	\pm	0.66	 $ & $ 	118.3	 $ & $	0.7	\sigma	 $  \\
$\delta C_{Q_{2}}^{\mu} $  	& $ 	0.00	\pm	0.02	 $ & $ 	118.7	 $ & $	0.3	\sigma	 $  \\
$\delta C_{Q_{2}}^{e} $    	& $ 	-0.77	\pm	0.65	 $ & $ 	118.4	 $ & $	0.6	\sigma	 $  \\
\hline
\end{tabular}
}  
\hspace*{1.5cm}
\scalebox{0.90}{
\begin{tabular}{|l|r|r|c|}
\hline 
 \multicolumn{4}{|c|}{All observables  ($\chi^2_{\rm SM}=118.8$)} \\ \hline
                          & b.f. value & $\chi^2_{\rm min}$ & ${\rm Pull}_{\rm SM}$  \\ 
\hline \hline
$\delta C_{9} $          	& $ 	-1.03	\pm	0.20	 $ & $ 	99.9	 $ & $	4.3	\sigma	 $  \\
$\delta C_{9}^{\prime}  $       & $ 	0.17	\pm	0.28	 $ & $ 	118.4	 $ & $	0.6	\sigma	 $  \\
$\delta C_{9}^{\mu} $      	& $ 	-1.11	\pm	0.17	 $ & $ 	85.1	 $ & $	5.8	\sigma	 $  \\
$\delta C_{9}^{e} $        	& $ 	1.22	\pm	0.33	 $ & $ 	103.8	 $ & $	3.9	\sigma	 $  \\
$\delta C_{9}^{\prime\,\mu} $   & $ 	0.04	\pm	0.19	 $ & $ 	118.7	 $ & $	0.3	\sigma	 $  \\
$\delta C_{9}^{\prime\,e} $     & $ 	0.08	\pm	0.30	 $ & $ 	118.7	 $ & $	0.3	\sigma	 $  \\
\hline
\end{tabular}
} \\ \vspace*{0.5cm} 
\hspace*{.5cm}
\scalebox{0.90}{
\begin{tabular}{|l|r|r|c|}
\hline 
 \multicolumn{4}{|c|}{All observables  ($\chi^2_{\rm SM}=118.8$)} \\ \hline
                          & b.f. value & $\chi^2_{\rm min}$ & ${\rm Pull}_{\rm SM}$  \\ 
\hline \hline
$\delta C_{10} $         	& $ 	0.21	\pm	0.25	 $ & $ 	118.0	 $ & $	0.9	\sigma	 $  \\
$\delta C_{10}^{\prime} $       & $ 	0.05	\pm	0.19	 $ & $ 	118.7	 $ & $	0.3	\sigma	 $  \\
$\delta C_{10}^{\mu} $     	& $ 	0.67	\pm	0.21	 $ & $ 	106.8	 $ & $	3.5	\sigma	 $  \\
$\delta C_{10}^{e} $       	& $ 	-1.06	\pm	0.28	 $ & $ 	103.2	 $ & $	3.9	\sigma	 $  \\
$\delta C_{10}^{\prime\,\mu} $  & $ 	0.04	\pm	0.16	 $ & $ 	118.7	 $ & $	0.3	\sigma	 $  \\
$\delta C_{10}^{\prime\,e} $    & $ 	-0.04	\pm	0.29	 $ & $ 	118.7	 $ & $	0.3	\sigma	 $  \\
\hline
\end{tabular}
} \hspace*{1.cm}
\scalebox{0.85}{
\begin{tabular}{|l|r|r|c|}
\hline 
 \multicolumn{4}{|c|}{All observables  ($\chi^2_{\rm SM}=118.8$)} \\ \hline
                          & b.f. value & $\chi^2_{\rm min}$ & ${\rm Pull}_{\rm SM}$  \\ 
\hline \hline
$\delta C_{\rm LL}^\mu$ \,  ($\delta C_{9}^{\mu}=-\delta C_{10}^{\mu}$)           	& $ 	-0.55	\pm	0.12	 $ & $ 	93.8	 $ & $	5.0	\sigma	 $  \\
$\delta C_{\rm LL}^e$ \,  ($\delta C_{9}^{e}=-\delta C_{10}^{e}$)			& $ 	0.60	\pm	0.17	 $ & $ 	103.4	 $ & $	3.9	\sigma	 $  \\
$\delta C_{\rm LR}^\mu$ \,  ($\delta C_{9}^{\mu}=+\delta C_{10}^{\mu}$)			& $ 	-0.35	\pm	0.17	 $ & $ 	115.1	 $ & $	2.0	\sigma	 $  \\
$\delta C_{\rm LR}^e$ \,  ($\delta C_{9}^{e}=+\delta C_{10}^{e}$)			& $ 	-1.86	\pm	0.32	 $ & $ 	103.3	 $ & $	3.9	\sigma	 $  \\
$\delta C_{\rm RR}^\mu$ \,  ($\delta C_{9}^{\mu \prime}=+\delta C_{10}^{\mu \prime}$)	& $ 	0.12	\pm	0.20	 $ & $ 	118.4	 $ & $	0.6	\sigma	 $  \\
$\delta C_{\rm RR}^e$ \,  ($\delta C_{9}^{e \prime}=+\delta C_{10}^{e \prime}$)		& $ 	2.13	\pm	0.33	 $ & $ 	103.2	 $ & $	3.9	\sigma	 $  \\
$\delta C_{\rm RL}^\mu$ \,  ($\delta C_{9}^{\mu \prime}=-\delta C_{10}^{\mu \prime}$)	& $ 	-0.01	\pm	0.09	 $ & $ 	118.8	 $ & $	0.1	\sigma	 $  \\
$\delta C_{\rm RL}^e$ \,  ($\delta C_{9}^{e \prime}=-\delta C_{10}^{e \prime}$)		& $ 	0.03	\pm	0.14	 $ & $ 	118.7	 $ & $	0.3	\sigma	 $  \\
\hline
\end{tabular}
}
\caption{Best fit values and errors in the one operator fits to all the relevant data on $b \to s$ transitions, assuming 10\% error for the power corrections. 
\label{tab:NPfit_1op}} 
\end{center} 
\end{table}

We first consider fits to one single Wilson coefficient. In Table~\ref{tab:NPfit_1op} the one-dimensional fit results are given for the Wilson coefficients $C_7, C_8, C_9^{\ell}, C_{10}^{\ell}, C_{Q_1}^\ell, C_{Q_2}^\ell$ as well as for the $O_{XY}^i$ basis which is well motivated in several NP models, where $X$ indicates the chirality of the quark current and $i$ and $Y $ stand for the flavour index and  chirality of the lepton current, respectively (see e.g. Ref.~\cite{Hurth:2014vma}).
NP  contributions to the primed Wilson coefficients, as well as $C_7,C_{Q_{1}},C_{Q_2}$ are disfavoured in the fit\footnote{For the $B\to K^* \mu^+ \mu^-$ observables we have used the LHCb results~\cite{Aaij:2015oid} within the most likelihood method where the scalar and tensorial Wilson coefficients are assumed to be zero. 
While alternatively, the data with the method of moments could be used, at present the experimental errors are very large.}, 
the same is also true for the axial-vector coefficient $C_{10}$ when lepton flavour universality is assumed.
In all favoured scenarios whenever lepton flavour universality violation is allowed, the fit is improved which is due to the tensions in $R_{K^{(*)}}$ measurements.
The most favoured scenario in the one-dimensional fit is when there is NP in $C_9^{\mu}$ with a significance of $5.8\sigma$. The scenario with NP in $C_{LL}^\mu$ has also an equally large significance of $5.8\sigma$.

\begin{figure}[th!]
\centering
\includegraphics[width=0.45\textwidth]{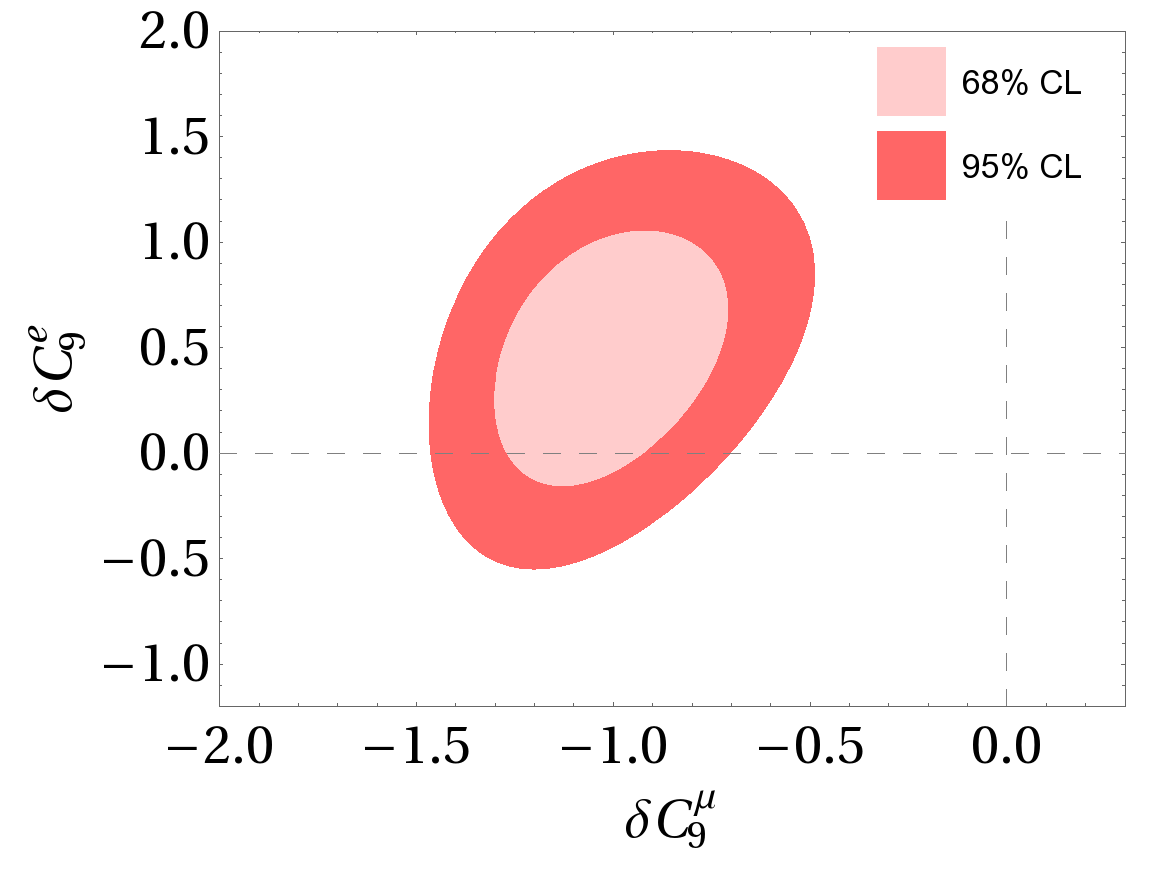}
\includegraphics[width=0.45\textwidth]{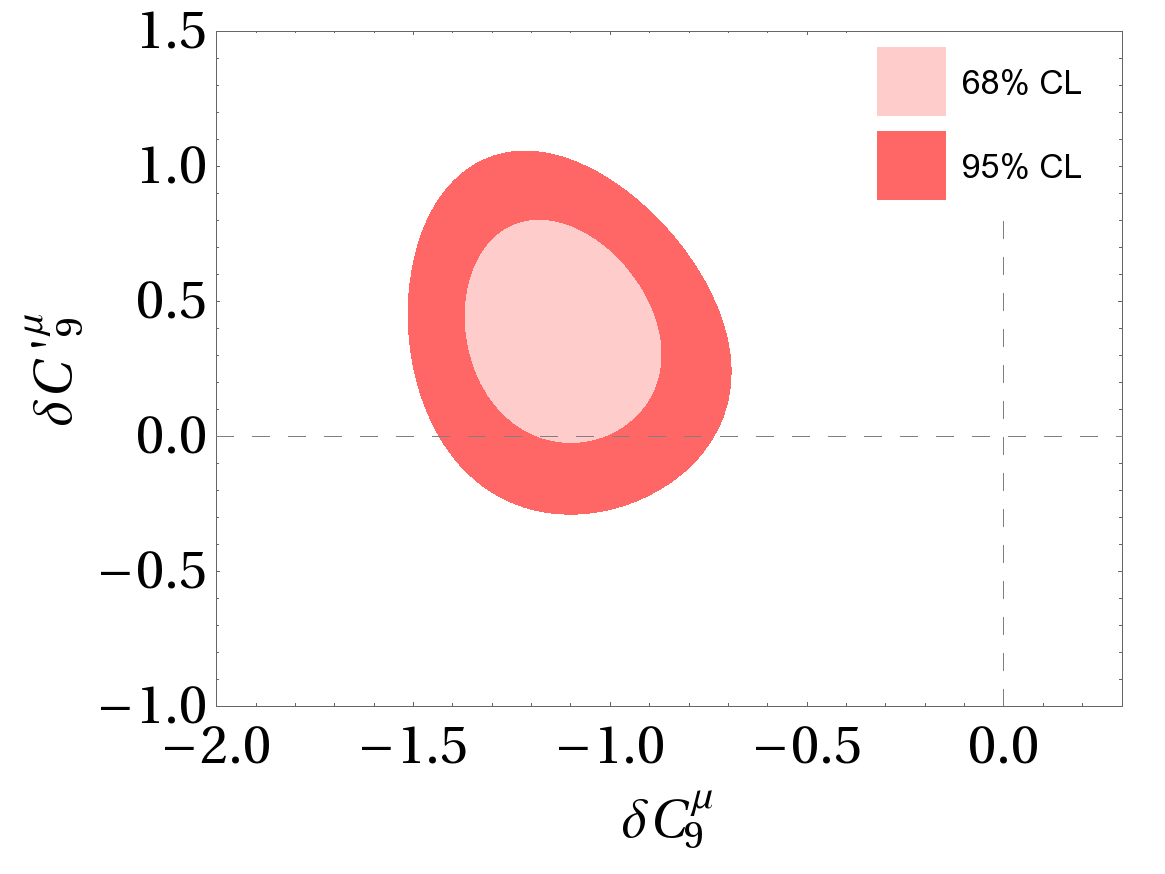}
\includegraphics[width=0.45\textwidth]{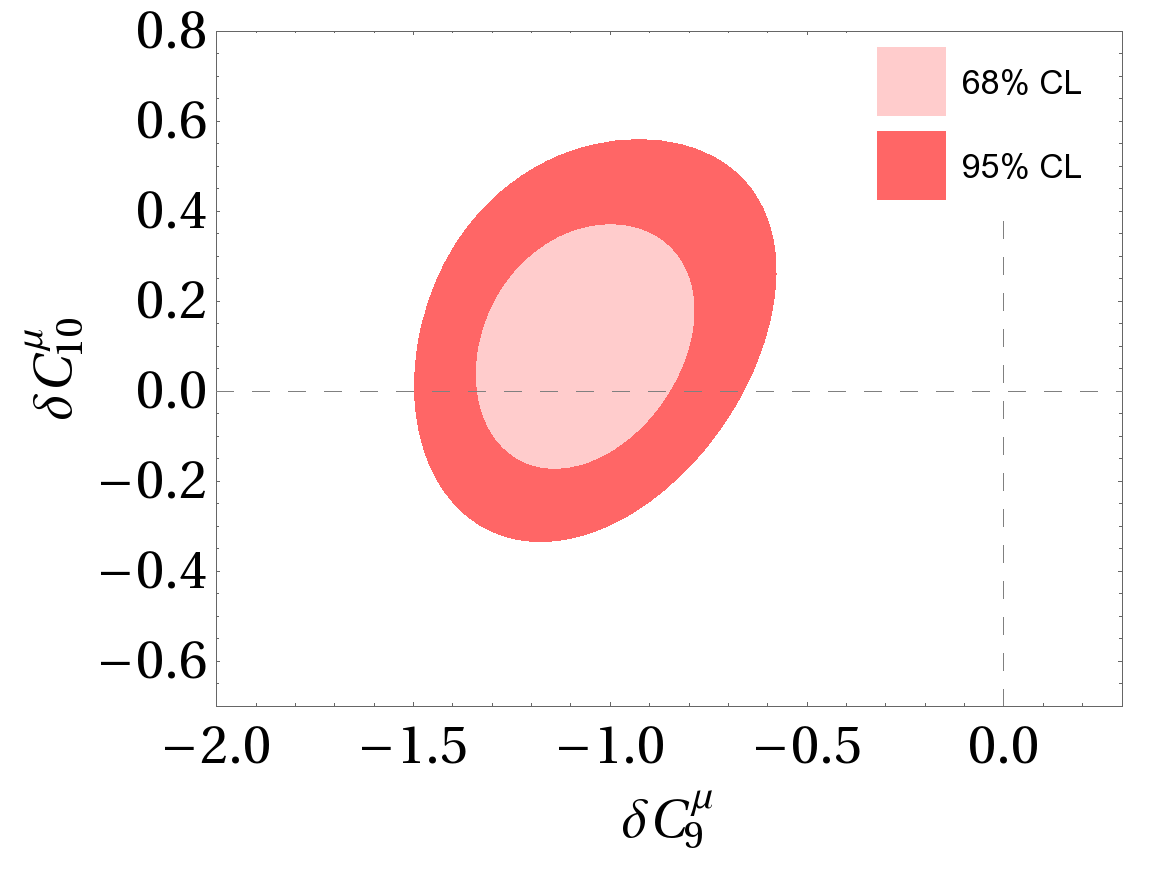}
\includegraphics[width=0.45\textwidth]{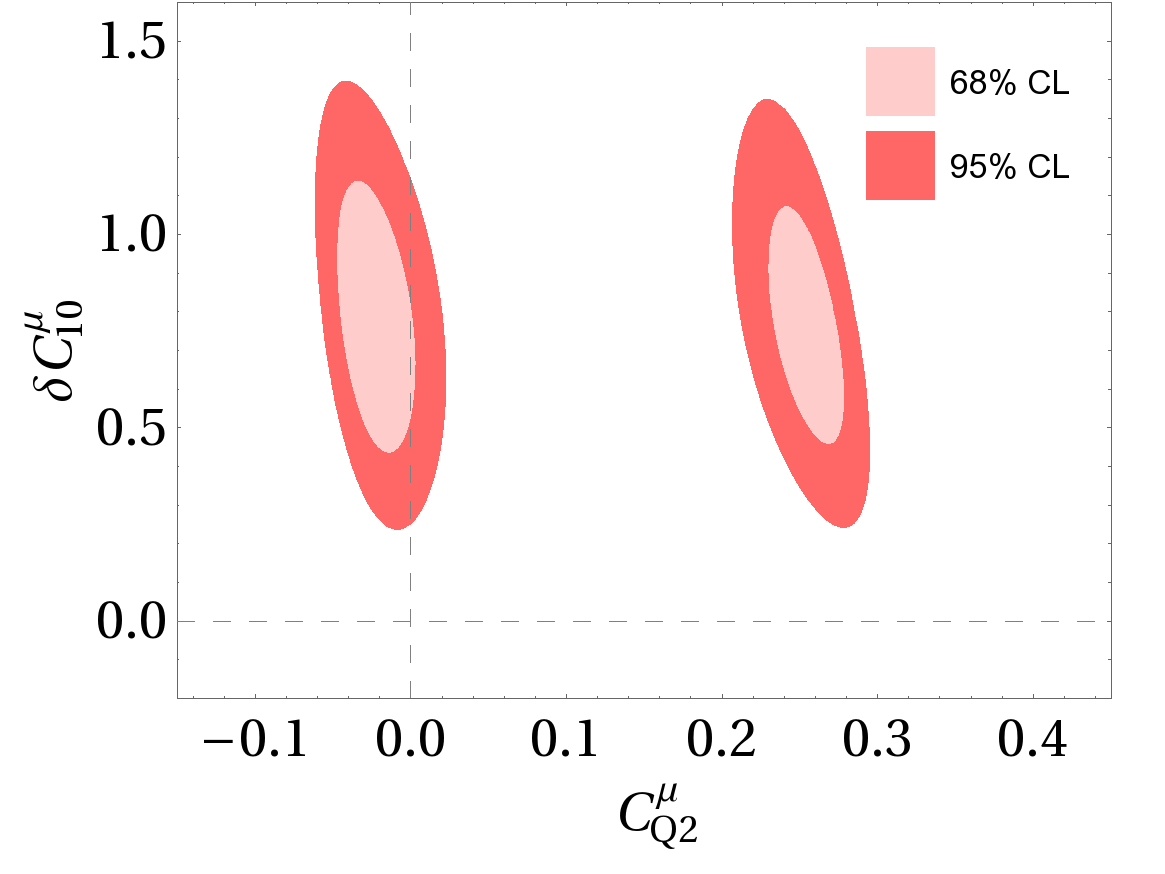}
\includegraphics[width=0.45\textwidth]{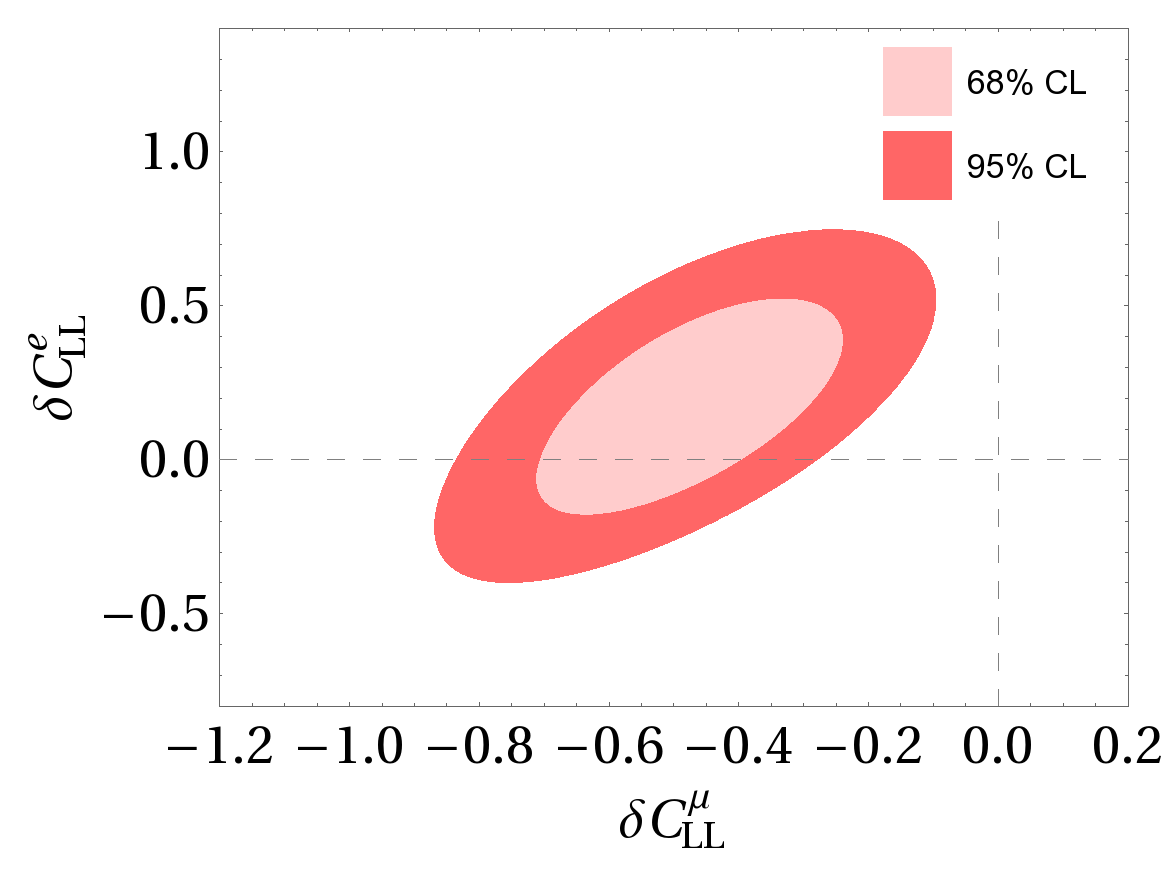}
\includegraphics[width=0.45\textwidth]{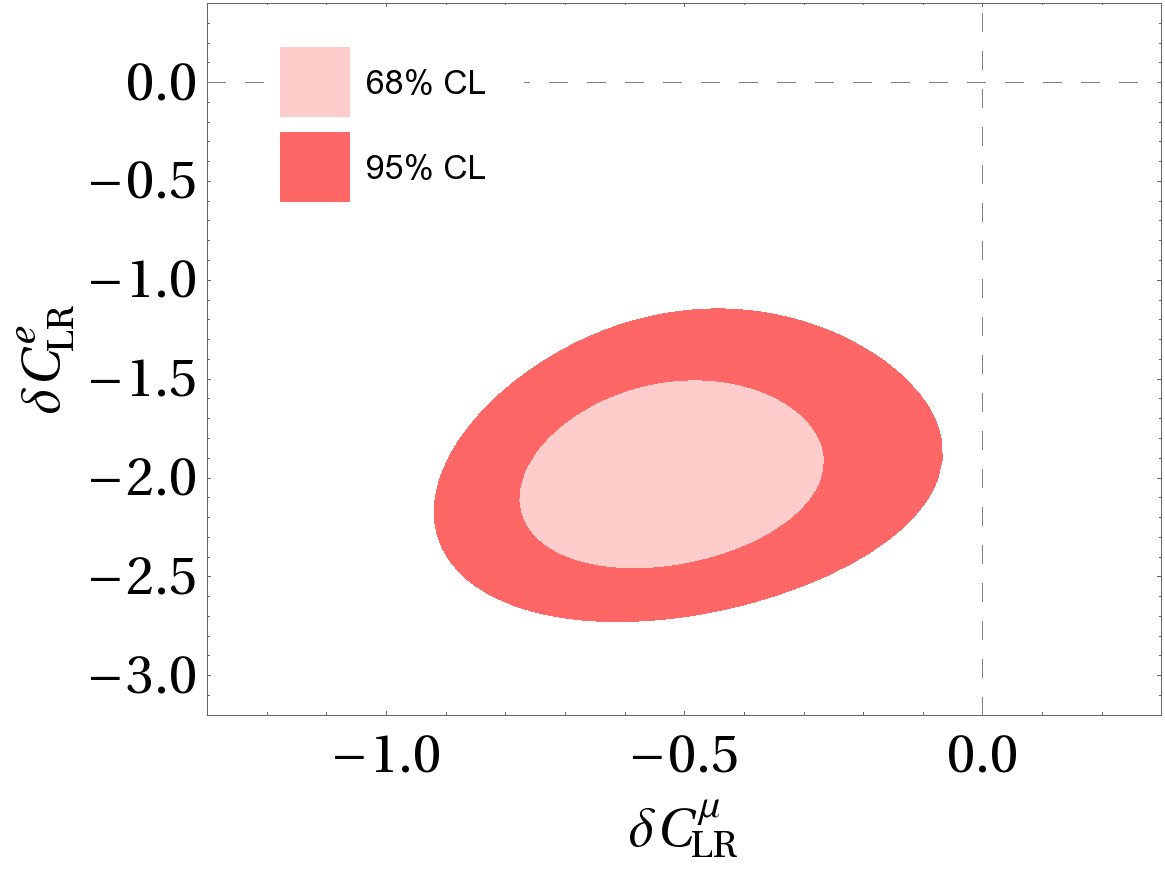}
\caption{Two operator global fits where the (light) red contour in the plots corresponds to the (68) 95\% C.L. regions, assuming 10\% error for the power corrections.
\label{Fig:NP_2op}}
\end{figure}

We now turn to two dimensional fits. We have performed 6 different fits, and their significance as well as the parameters of the best fit points are given in Table~\ref{tab:allobs_2D}. A graphical representation showing the 68 and 95\% C.L. contours is also provided in Fig.~\ref{Fig:NP_2op}.
The fit corresponding to $\delta C_{10}^{\mu},C_{Q_2}^{\mu}$ illustrates our discussion on the branching ratio of $B_s \to \mu^+ \mu^-$, showing that the best fit point corresponds to $C_{Q_2}^{\mu}\sim0$, but $C_{10}^{\mu}$ can receive a rather large deviation from its SM value. Yet, the pull with the SM is only 3.3$\sigma$. Scenarios with $C_{LL}$ and $C_{LR}$ improve the fits, leading to significances of more than $4\sigma$. 
The most favoured scenarios are for the case where there is NP in $C_{9}^{\mu}$ in combination with $\delta C_{9}^{\prime \mu}$, $\delta C_{9}^e$ or $\delta C_{10}^\mu$, with significances of $\sim5.5\sigma$ in very good agreement with the results of Table II in Ref.~\cite{Capdevila:2017bsm}.

\begin{table}[th!]
\begin{center}
\setlength\extrarowheight{3pt}
\scalebox{0.8}{
\begin{tabular}{|l|c|c|c|}
\hline 
 \multicolumn{4}{|c|}{All observables  ($\chi^2_{\rm SM}=118.8$)} \\ \hline
                          & b.f. value & $\chi^2_{\rm min}$ & ${\rm Pull}_{\rm SM}$ \\ 
\hline \hline
$\{ \delta C_{9}^{\mu} , \delta C_{9}^{\prime \mu} \}$	& $\{ 	-1.14	 \pm 	0.16	 \;,\; 	0.39	 \pm 	0.27	 \}$ & $ 	83.0	 $ & $ 5.64\sigma$   \\
$\{ \delta C_{9}^{\mu} , \delta C_{9}^{e} \}$		& $\{ 	-1.03	 \pm 	0.19	 \;,\; 	0.45	 \pm 	0.40	 \}$ & $ 	83.9	 $ & $ 5.56\sigma$   \\
$\{ \delta C_{9}^{\mu} , \delta C_{10}^{\mu} \}$	& $\{ 	-1.08	 \pm 	0.18	 \;,\; 	0.09	 \pm 	0.18	 \}$ & $ 	84.8	 $ & $ 5.48\sigma$   \\
$\{ \delta C_{10}^{\mu} , C_{Q_{2}}^{\mu} \}$	& $\{ 	0.78	 \pm 	0.23	 \;,\; 	-0.02	 \pm 	0.02	 \}$ & $ 	104.8	 $ & $ 3.32\sigma$   \\
$\{ \delta C_{LL}^{\mu} , \delta C_{LL}^{e} \}$		& $\{ 	-0.48	 \pm 	0.16	 \;,\; 	0.17	 \pm 	0.23	 \}$ & $ 	93.3	 $ & $ 4.68\sigma$   \\
$\{ \delta C_{LR}^{\mu} , \delta C_{LR}^{e} \}$		& $\{ 	-0.54	 \pm 	0.17	 \;,\; 	-2.01	 \pm 	0.31	 \}$ & $ 	95.5	 $ & $ 4.45\sigma$   \\
\hline
\end{tabular}
}
\caption{Best fit values and errors in the two operator global fits, assuming 10\% error for the power corrections.
\label{tab:allobs_2D}} 
\end{center} 
\end{table}

We now expand the fits of the Wilson coefficients to $6, 10$ and $20$ dimensions. The results of the fits including $C_9^\mu$ are given in Table.~\ref{tab:NPimprovements}. The improvement column corresponds to the improvement in comparison to the previous set of Wilson coefficients, obtained using the Wilks' theorem. Additional fit results can be found in appendix~\ref{sec:additionalfits}, for $\{C_{10}, C_{Q_1}, C_{Q_2} \}$ (Table~\ref{tab:allobs_3D_C10Q1Q2}), $\{C_{10}^{e,\mu}, C_{Q_1}^{e,\mu}, C_{Q_2}^{e,\mu} \}$ (Table~\ref{tab:allobs_6D_C10Q1Q2primes}), $\{C_7,C_8,C_9^{(e,\mu)},C_{10}^{(e,\mu)}\}$ (Table~\ref{tab:allobs_6D_C78910}) and $\{C_7,C_8,C_9^{(e,\mu)},C_{10}^{(e,\mu)},C_{Q_1}^{e,\mu}, C_{Q_2}^{e,\mu} \}$ (Table~\ref{tab:allobs_10D_C78910C12}), including the best fit point values.

The pull with the SM decreases with the number of Wilson coefficients. The reason is due to the fact that increasing the number of Wilson coefficients raises the number of degrees of freedom. In absence of real improvement in the fit, i.e. a strong decrease in the best fit point $\chi^2$, the increase of number of degrees of freedom will result in a reduced pull with the SM. 
This is confirmed by the improvement test, which reveals that adding Wilson coefficients to the ``$C_9^\mu$ only'' set does not bring any significant improvement.
This result is in agreement with several recent fits with similar sets of observables (see e.g. Refs.~\cite{Capdevila:2017bsm,DAmico:2017mtc}) where the global analysis of $b\to s \ell^+ \ell^-$ data indicates preference for NP scenarios with modified $C_9^\mu$ with a significance of larger than $5\sigma$.

\begin{table}[!th]
\begin{center}
\scalebox{1.}{
\begin{tabular}{|c|c|c|c|c|}
\hline
Set of WC & Nr. parameters & $\chi^2_{\rm min}$ & Pull$_{\rm SM}$ & Improvement\\
\hline\hline
SM & 0 & 118.8 & - & -\\
\hline
$C_9^{\mu}$ & 1 & 85.1 & $5.8\sigma$ & $5.8\sigma$\\
\hline
$C_9^{(e,\mu)}$ & 2 & 83.9 & $5.6\sigma$ & $1.1\sigma$\\
\hline
$C_7,C_8,C_9^{(e,\mu)},C_{10}^{(e,\mu)}$ & 6 & 81.2 & $4.8\sigma$ & $0.5\sigma$\\
\hline
All non-primed WC & 10 (8) & 81.0 & $4.1\,(4.5)\sigma$ & $0.0\,(0.1)\sigma$\\
\hline
All WC (incl. primed) & 20 (16) & 70.2 & $3.6\,(4.1)\sigma$ & $0.9\,(1.2)\sigma$\\
\hline
\end{tabular}
}
\caption{The $\chi^2_{\rm min}$ values when varying different Wilson coefficients. 
In the last column the significance of the improvement of the fit compared to the scenario of the previous line is given, (assuming 10\% error for the power corrections). The numbers in the parenthesis correspond to removing $C_{Q_{1,2}}^{e \, (\prime)}$ from the number of degrees of freedom. 
\label{tab:NPimprovements}}
\end{center} 
\end{table}

\begin{table}[th!]
\begin{center}
\setlength\extrarowheight{3pt}
\scalebox{0.85}{
\begin{tabular}{|c|c|c|c|}
\hline 
  \multicolumn{4}{|c|}{All observables  with $\chi^2_{\rm SM}=118.8$} \\ 
  \multicolumn{4}{|c|}{($\chi^2_{\rm min}=70.2;\; {\rm Pull}_{\rm SM}=3.5\,(4.1)\sigma$)} \\ 
\hline \hline
\multicolumn{2}{|c}{$\delta C_7$} &  \multicolumn{2}{|c|}{$\delta C_8$}\\ 
\multicolumn{2}{|c}{$-0.01  \pm 0.05 $} & \multicolumn{2}{|c|}{$ 0.89 \pm 0.81 $}\\ 
\hline 
\multicolumn{2}{|c}{$\delta C_7^\prime$} &  \multicolumn{2}{|c|}{$\delta C_8^\prime$}\\ 
\multicolumn{2}{|c}{$ 0.01  \pm 0.03 $} & \multicolumn{2}{|c|}{$ -1.70 \pm 0.46 $}\\ 
\hline \hline
$\delta C_{9}^{\mu}$ & $\delta C_{9}^{e}$ & $\delta C_{10}^{\mu}$ & $\delta C_{10}^{e}$ \\
$ -1.40 \pm 0.26 $ & $ -4.02  \pm 5.58 $ & $-0.07 \pm 0.28 $ & $ 1.32  \pm 2.02 $  \\
\hline 
$\delta C_{9}^{\prime \mu}$ & $\delta C_{9}^{\prime e}$ & $\delta C_{10}^{\prime \mu}$ & $\delta C_{10}^{\prime e}$ \\
$ 0.23 \pm 0.65 $ & $-1.10  \pm 5.98 $ & $ -0.16 \pm 0.38 $ & $ 2.70  \pm 2 $ \\
\hline \hline
$C_{Q_{1}}^{\mu}$ & $C_{Q_{1}}^{e}$ & $C_{Q_{2}}^{\mu}$ & $C_{Q_{2}}^{e}$ \\ 
$-0.13  \pm 1.86 $ & undetermined & $-0.05  \pm 0.58 $ & undetermined \\
\hline 
$C_{Q_{1}}^{\prime \mu}$ & $C_{Q_{1}}^{\prime e}$ & $C_{Q_{2}}^{\prime \mu}$ & $C_{Q_{2}}^{\prime e}$ \\ 
$ 0.01  \pm 1.87 $ & undetermined & $-0.18  \pm 0.62 $ & undetermined \\
\hline
\end{tabular}
}
\caption{Best fit values for the 20 operator global fit to the $b \to s$ data, assuming 10\% error for the power corrections.
\label{tab:allobs_20D_C78910C12primes}} 
\end{center} 
\end{table}

As a UV-complete NP model is likely to incorporate several new particles affecting all the Wilson coefficients, we give in Table~\ref{tab:allobs_20D_C78910C12primes} the best fit values when varying all the 20 Wilson coefficients.
Several Wilson coefficients have loose constraints which is due to the large number of free parameters compared to 
the numbers of observables and also the lack of observables with sufficient sensitivity to those Wilson coefficients. 
The best fit values indicate potentially large contributions to the electron Wilson coefficients ($C_{9,10,{Q_1}}^{e(\prime)}$) which is interesting as the few measurements on purely electron observables are much more SM-like than their muon counterparts and is mostly driven by the flavour violating observables $R_{K^{(*)}}$.
However, the large contributions are not statistically significant as there are many more muon than electron observables in the global fit. Specifically, the favoured large contribution in the electron scalar coefficient is due to the absence of constraining experimental results on the $B_s \to e^+ e^-$  decay which would be sensitive enough to the scalar and pseudoscalar Wilson coefficients, the latter remaining currently completely undetermined in the 20-dimensional fit. It can be noted that $C_7$ and $C_7^{\prime}$ are severely constrained with very small errors, revealing the compatibility between the constraints. $C_8^{(\prime)}$ is much less constrained, as there are less observables sensitive to $C_8$ in the fit. In addition, the muon scalar and pseudoscalar contributions can only have very small values. The best fit value of $C_9^\mu$ is even smaller than for the one and two-dimensional fits, with 35\% reduction compared to its SM value.

A comment about the number of degrees of freedom is in order here. As can be seen from Table~\ref{tab:NPimprovements}, $C_{Q_{1,2}}^{e \, (\prime)}$ are ``undetermined'' due to their very large uncertainties. We checked explicitly how the variation of order one in each Wilson coefficient affects the $\chi^2$, which confirmed that the four $C_{Q_{1,2}}^{e \, (\prime)}$ coefficients have a negligible impact on the fit, i.e. for each coefficient $|\delta C_i| \sim 1$ implies $|\delta \chi^2| < 1$. Therefore, one can define an effective number of degrees of freedom in which the insensitive coefficients are not counted. The results are shown in parentheses in Table ~\ref{tab:NPimprovements}.

Finally, as a result of the full fit including all the relevant Wilson coefficients, we obtain a total pull of 4.1$\sigma$ with the SM hypothesis (assuming 10\% error for the power corrections).

\section{Conclusions}\label{sec:conclusions}
Recent experimental measurements have shown tensions in some of the $b \to s$ transitions.
The most persistent tension which has been confirmed by several experiments is the anomaly 
in the angular observable $P_5^\prime$ of the $B\to K^* \mu^+ \mu^-$ decay. 
This decay, however, receives long-distance hadronic contributions that are difficult to calculate
and consequently makes the SM predictions  somewhat questionable.
Hence the significance of the observed tensions is quite dependent on how the non-factorisable contributions are estimated.
In this paper we explored the various state-of-the-art methods for implementing  the power corrections and 
demonstrated that while the various implementations of the unknown corrections offer different SM predictions
and uncertainties, in all these cases, in the critical bin where the $P_5^\prime$ anomaly is observed,
the predictions roughly converge giving prominence to the observed tensions.

Alternatively, instead of making assumptions on the size of the power corrections or using methods which include these contributions (and introduce in some cases non-transparent systematic uncertainties and correlations as we have shown) one can assume a general parameterisation for the power corrections and fit the unknown parameters of the ansatz to the $B \to K^* \mu^+ \mu^-$ data.
In this work, in addition to the $B \to K^* \mu^+ \mu^-$ observables
we have included data on BR($B \to K^* \gamma$) which requires the ansatz for the power corrections
to have the correct end-point behaviour as the virtual photon 
(which decays into the dimuon in $B \to K^* \mu^+ \mu^-$) becomes on-shell.
The ansatz employed in this paper is the most general parameterisation (up to higher $q^2$ terms) which respects the analyticity structure of the amplitudes and guarantees that  the longitudinal amplitude disappears as $q^2 \to 0$.

Employing this model-independent ansatz we examined whether NP  contribution 
to (real or complex) $C_{9}$ and $C_7$ Wilson coefficients (with 1-4 free parameters) 
is the favoured explanation for the anomalies or underestimated hadronic effects (modeled with 18 free parameters).
A statistical comparison indicates that there is no significant preference in adding 14-17 parameters compared to the NP explanation.
This is partly due to the experimental results not being constraining enough so that the 18 parameters of 
the power corrections are mostly consistent with zero and also since possible preference for a 
large $q^2$-dependence might be masked due to the $q^2$ smearing within the current ranges of the bins.
Furthermore, when employing only CP-averaged flavour universal observables, due to the embedding of the NP contributions in the hadronic effects the latter cannot be ruled out in favour of the former while the opposite is possible. 
Therefore, whilst still the most favoured scenario is having real NP contributions in $C_9$, the picture remains inconclusive and more precise data with finer binning will be crucial in clarifying the situation, especially on CP-asymmetric observables which can differentiate the weak and strong phases emerging from NP and hadronic contributions, respectively.
Thus, the Wilks' test established in this paper will be a very important tool to analyse the forthcoming  
$B\to K^* \mu^+ \mu^-$ data.

Furthermore, we presented here for the first time a global fit to the present $b \to s$ data using all effective parameters and fixed the NP significance. We found a total pull of 4.1$\sigma$ with the SM hypothesis (assuming 10\% error for the power corrections). We also showed that while BR($B_s \to \mu^+ \mu^-$) is very effective in constraining the scalar and pseudoscalar operators,
the relevant Wilson coefficients cannot be neglected by only assuming this single observable and a global fit to all the $b \to s$ data is required where all relevant Wilson coefficients can simultaneously receive NP contributions. Although, the various 1, 2, 6, 10 and 20 dimensional fits when varying different Wilson coefficients do not indicate any preference for NP beyond $C_9$ using the present data, yet a large number of Wilson coefficients are very loosely bound or completely undetermined in the case of electron scalar and pseudoscalar operators. 
This is interesting since especially with the indication of lepton flavour universality violation from the $R_K$ and $R_{K^*}$ ratios, there is motivation to investigate the electron and muon sectors separately for the scalar and pseudoscalar operators.

\section*{Acknowledgement} 
The authors are grateful to Christoph Bobeth for suggesting to consider the angular observable $F_H(B^+ \to K^+ \mu^+ \mu^-)$ in the global fit, and to Danny van Dyk, Alexander Khodjamirian and Luca Silvestrini for clarifying communications.
TH thanks the CERN theory group for its hospitality during his regular visits to CERN where part of this work was written.

\clearpage

\appendix

\section{The $q^2$-dependence of $H_V(\lambda)$ for $\lambda=\pm$ and $\lambda=0$}\label{sec:q2behaviour}

We show in the following that the effect of NP contributions to $B \to K^* \ell^+ \ell^-$ observables from $C_7$ and $C_9$ 
can be embedded in the most general ansatz of the hadronic contributions. Thus  it is possible to make a statistical comparison of a hadronic fit and a NP fit of $C_9$ (and $C_7$) to the $B\to K^* \mu^+ \mu^-$ data.
We note here that the form factors $\tilde{V}_\lambda,\tilde{T}_\lambda$ appearing in $H_V(\lambda)$ (Eq.~(\ref{eq:HV})) have different $q^2$-behaviours for $\lambda=\pm$ and  $\lambda=0$.

\subsection{$H_V(\lambda=\pm)$}\label{sec:lambdaPMbehaviour}
\begin{figure}[h!]
\centering
\includegraphics[width=0.48\textwidth]{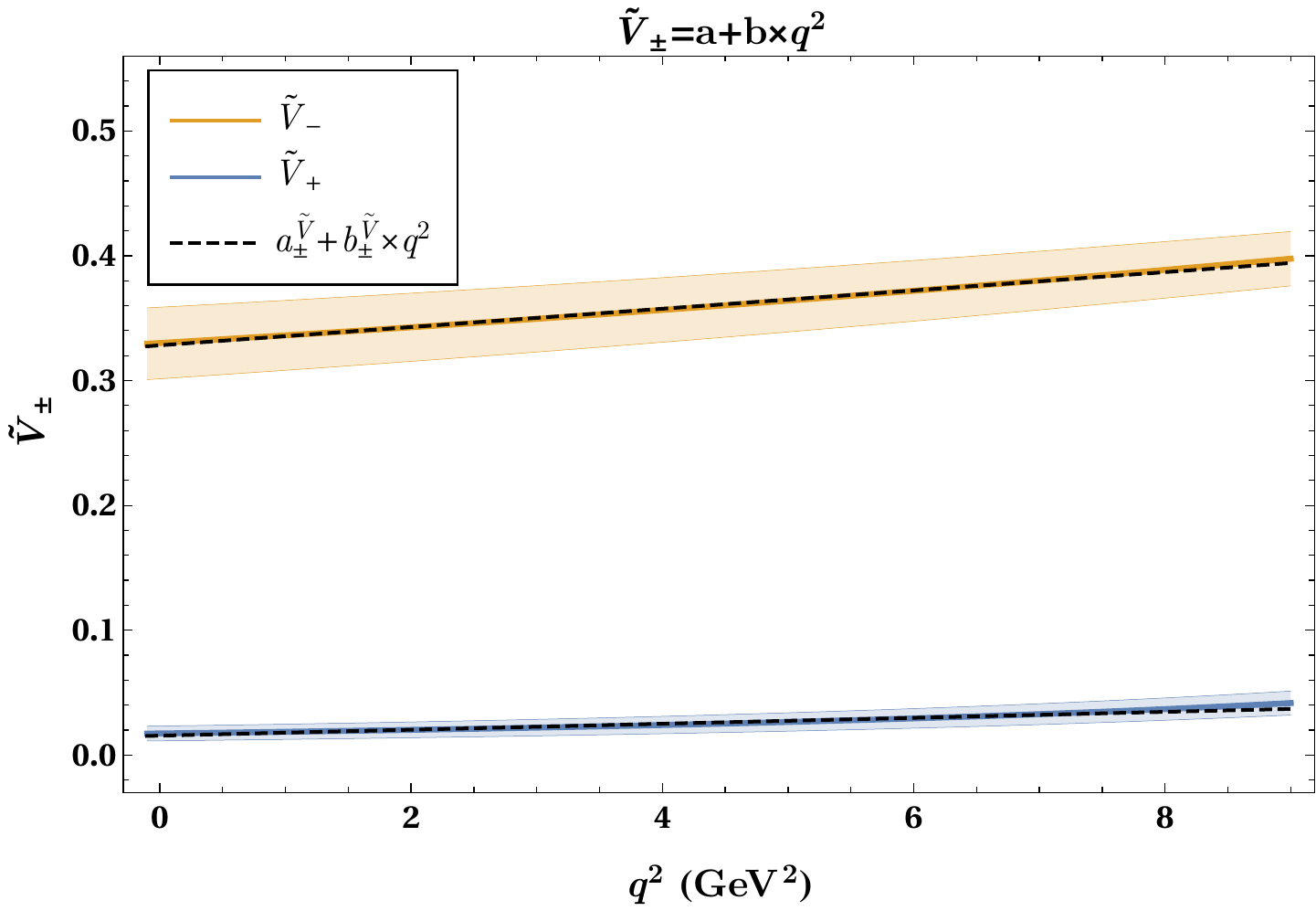}
\includegraphics[width=0.48\textwidth]{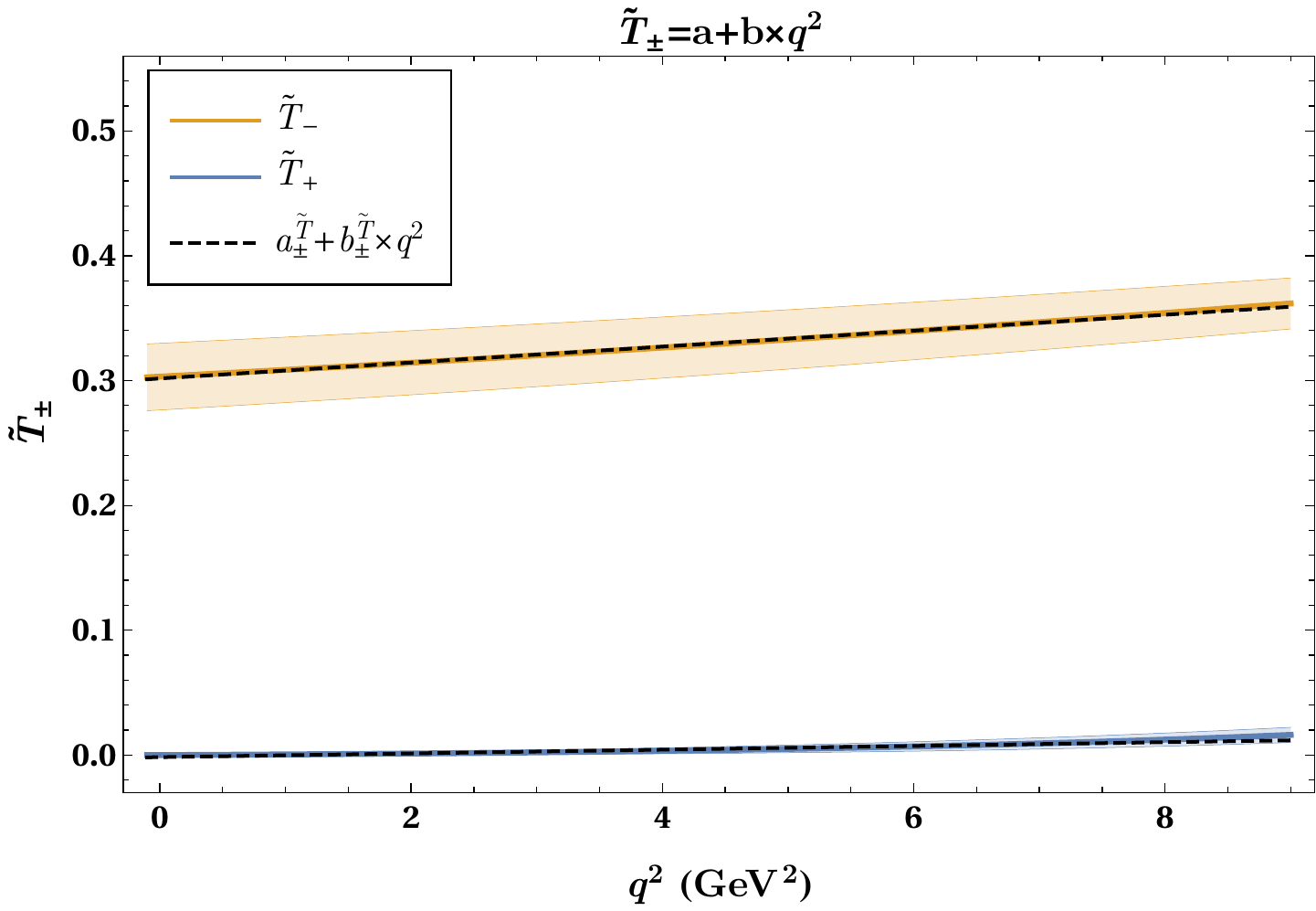}
\caption{Form factors $\tilde{V}_\pm$ and $\tilde{T}_\pm$, where the solid lines correspond
to the analytical expression and the dashed lines represent the expanded function. 
The helicity form factor error bands are calculated from the  uncertainties and correlations  of the ``LCSR + Lattice'' fit results for the traditional form factors $V,A_{1,12}$ and $T_{1,2,23}$ of Ref.~\cite{Straub:2015ica}. 
\label{Fig:tildeFpmExpansion}}
\end{figure}

The helicity form factors $\tilde{V}_{\lambda=\pm}$ and $\tilde{T}_{\lambda=\pm}$ are written as
\begin{align}
\tilde{V}_{\pm}\left( q^{2}\right) &=\frac{1}{2} \bigg[ \Big( 1 + \frac{m_{K^*}}{m_B} \Big) A_1\left( q^{2}\right) \mp \frac{\lambda^{1/2}}{m_B(m_B + m_{K^*})} V\left( q^{2}\right) \bigg]\,, \nonumber\\
\tilde{T}_{\pm}\left( q^{2}\right) &= \frac{m_B^2 - m_{K^*}^2}{2m_B^2}T_2\left( q^{2}\right) \mp \frac{\lambda^{1/2}}{2m_B^2}T_1\left( q^{2}\right)\,, \nonumber
\end{align}
with $\lambda= m_B^4  + m_{K^*}^4 + q^4 - 2 (m_B^2 m_{K^*}^2+ m_{K^*}^2 q^2  + m_B^2 q^2)$. 
Since $V(q^2),A_1(q^2),T_1(q^2)$ and $T_2(q^2)$ are all well-behaved functions of $q^2$ (e.g. see Fig. 2 in Ref.~\cite{Straub:2015ica}), the helicity amplitudes 
$\tilde{V}_\pm$ and $\tilde{T}_\pm$ can be  described in terms of polynomials in $q^2$ (see Fig.~\ref{Fig:tildeFpmExpansion})
\begin{align}\label{eq:tildeFpmExpansion}
 \tilde{V}_\pm&= a^{\tilde{V}}_\pm +q^2\,b^{\tilde{V}}_\pm\,,\nonumber \\ 
 \tilde{T}_\pm&= a^{\tilde{T}}_\pm +q^2\,b^{\tilde{T}}_\pm\,,
\end{align}
where $a_\pm^{\tilde{V},\tilde{T}},b_\pm^{\tilde{V},\tilde{T}}$ 
are determined by expanding the form factors $\tilde{V}_\pm$ and $\tilde{T}_\pm$.

With the above expansion for the helicity form factors in Eq.~(\ref{eq:tildeFpmExpansion}), the effect of $\delta C_9^{\rm NP}$ and $\delta C_7^{\rm NP}$ in $H_V(\lambda=\pm)$ can be written as
\begin{align}\label{eq:NPLambdaPM}
 \delta H_V^{C_9^{\rm NP}}\!(\lambda=\pm) &= -i N^\prime \, \delta C_{9}^{\rm NP}\, \left(  a^{\tilde{V}}_\pm + q^2\, b^{\tilde{V}}_\pm   \right)\,,  \nonumber \\ 
 \delta H_V^{C_7^{\rm NP}}\!(\lambda=\pm) &= -i N^\prime 2\,\hat{m}_b\, m_B \, \delta C_{7}^{\rm NP}\, \left( \frac{1}{q^2}\, a^{\tilde{T}}_\pm +b^{\tilde{T}}_\pm \right)\,.
\end{align}

Employing the polynomial ansatz of Eq.~(\ref{eq:hlambda}), the effect of the power corrections is
\begin{align}\label{eq:haronicLambdaPM}
\delta H_V^{{\rm PC}}(\lambda=\pm) &= i\, N^\prime m_B^2\, 16 \pi^2 \left( \frac{1}{q^2}\, h_\lambda^{(0)} + h_\lambda^{(1)} + q^2\,h_\lambda^{(2)} \right)\,,
\end{align}
which is compatible with the form factor terms in $H_V(\lambda=\pm)$ and will not disrupt the analyticity structure of the amplitude.

Considering Eq.~(\ref{eq:NPLambdaPM}) and Eq.~(\ref{eq:haronicLambdaPM}),  NP  effect can clearly be embedded in the more general case of hadronic contributions.
Moreover, assuming  $b_\pm^{\tilde{V},\tilde{T}}$ in the Taylor expansions of the form factors $\tilde{T}_\pm,\tilde{V}_\pm$ to be zero, 
the $\delta C_9$ contributions correspond to $h^{(1)}_\pm$ and the $\delta C_7$ contributions to $h^{(0)}_\pm$ terms of the power corrections.

\subsection{$H_V(\lambda=0)$}\label{sec:lambda0behaviour}
\begin{figure}[h!]
\centering
\includegraphics[width=0.48\textwidth]{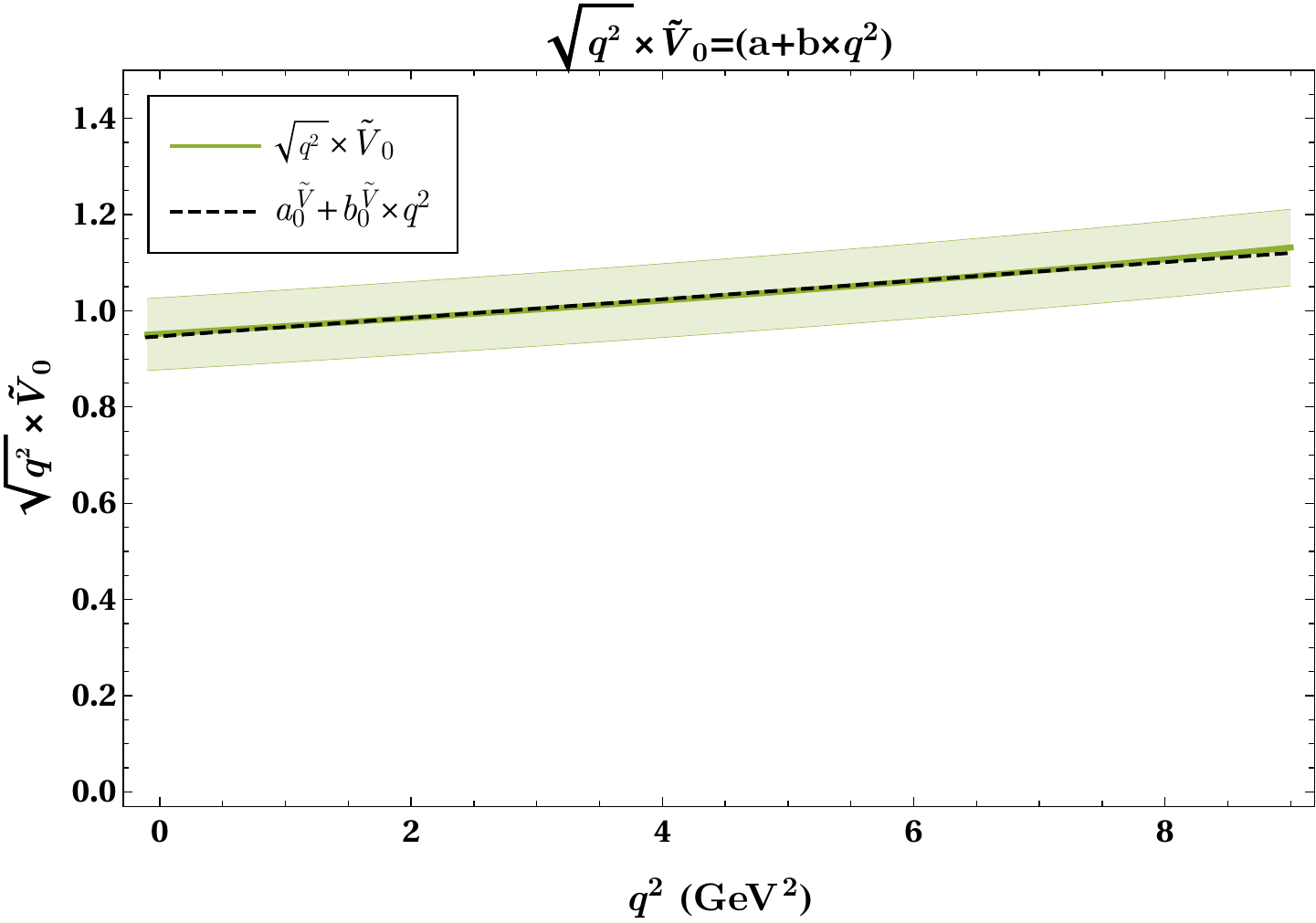}
\includegraphics[width=0.48\textwidth]{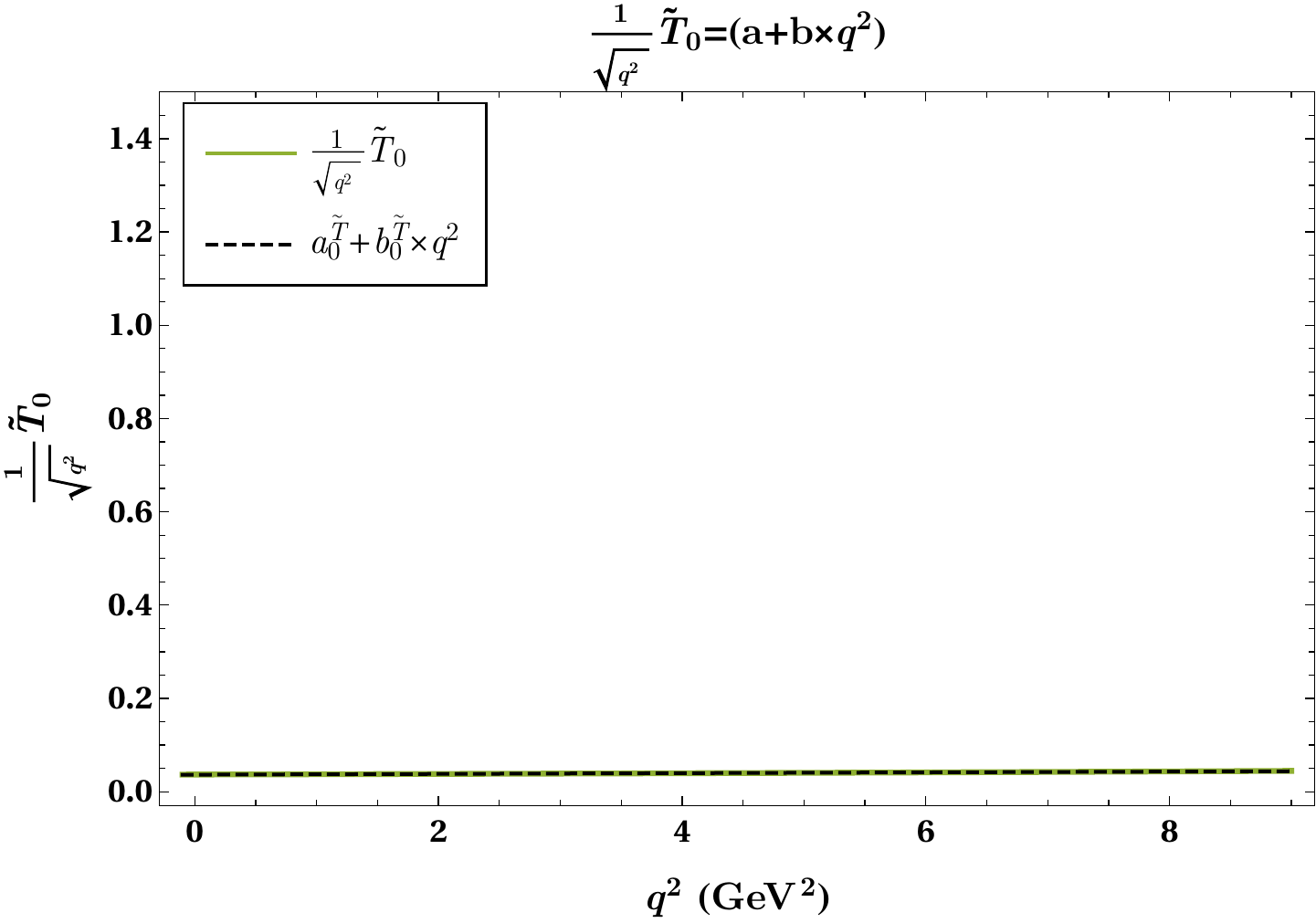}
\caption{Form factors $\sqrt{q^2}\times \tilde{V}_0$ and $(1/\sqrt{q^2})\times \tilde{T}_0$, where the solid lines show the
analytical expression and the dashed lines correspond to the expanded functions.\label{Fig:tildeF0Expansion}}
\end{figure}

The helicity form factors $\tilde{V}_{0}(q^2)$ and $\tilde{T}_{0}(q^2)$ are described in terms of $A_{12}(q^2)$ and $T_{23}(q^2)$ with an extra term of $1/\sqrt{q^2}$ and $\sqrt{q^2}$, respectively:
\begin{align}
\tilde{V}_0(q^2) = \frac{4m_{K^*}}{\sqrt{q^2}}A_{12}(q^2) && \text{and}  &&  \tilde{T}_0(q^2)=\frac{2\sqrt{q^2}m_{K^*}}{m_B(m_B + m_{K^*})}T_{23}(q^2),
\end{align}
where $A_{12}(q^2)$ and $T_{23}(q^2)$ 
are well-behaved functions of $q^2$ (see e.g. Fig. 2 in Ref.~\cite{Straub:2015ica}).
The helicity amplitudes 
$\tilde{V}_0$ and $\tilde{T}_0$ can then be described as a power expansion in $q^2$ in terms of
\begin{align}\label{eq:tildeF0Expansion}
 \tilde{V}_0= \frac{1}{\sqrt{q^2}}\left( a^{\tilde{V}}_0 +b^{\tilde{V}}_0 q^2 \right)  && \text{and}  &&   \tilde{T}_0= \sqrt{q^2} \left( a^{\tilde{T}}_0 +b^{\tilde{T}}_0 q^2 \right) \,,
\end{align}
where $a_0^{\tilde{V},\tilde{T}},b_0^{\tilde{V},\tilde{T}}$ 
are determined by expanding the form factors $\tilde{V}_0$ and $\tilde{T}_0$ (see Fig.~\ref{Fig:tildeF0Expansion}).

Considering the expansion in Eq.~(\ref{eq:tildeF0Expansion}) for the helicity form factors, the effect of $\delta C_9^{\rm NP}$ and $\delta C_7^{\rm NP}$ in $H_V(\lambda=\pm)$ can be written as
\begin{align}
 \delta H_V^{C_9^{\rm NP}}\!(\lambda=0) &= -i N^\prime \, \delta C_{9}^{\rm NP}\,\left[ \frac{1}{\sqrt{q^2}}\left(   a^{\tilde{V}}_0 + q^2 b^{\tilde{V}}_0  +  \right) \right] \,, \nonumber \\ 
 \delta H_V^{C_7^{\rm NP}}\!(\lambda=0) &= -i N^\prime 2\,\hat{m}_b\, m_B \, \delta C_{7}^{\rm NP}\, \left[ \frac{1}{\sqrt{q^2}}\left(   a^{\tilde{T}}_0 + q^2 b^{\tilde{T}}_0   \right) \right]\,.
\end{align}

Using the power expansion ansatz in Eq.~(\ref{eq:hzero}), the effect of the power corrections is
\begin{align}
\delta H_V^{{\rm PC}}(\lambda=0) &= i\, N^\prime m_B^2 \, 16 \pi^2 \left[\frac{1}{\sqrt{q^2}} \left(  h_0^{(0)} +q^2\, h_0^{(1)} + q^4\,h_0^{(2)} \right) \right]\,,
\end{align}
which results in terms that are compatible with the form factor terms in $H_V(\lambda=0)$ and will not disrupt the analyticity structure of the amplitude.
And the embedding of the NP effects in the hadronic contributions remains valid.
However, for $\lambda=0$ when assuming $b_0^{\tilde{V},\tilde{T}}$ in the Taylor expansions of the form factors $\tilde{V}_0,\tilde{T}_0$ to be zero, $C_9$ and $C_7$ both correspond to $h^{(0)}_0$.

Considering $H_V(\lambda=0)$ it might seem that the longitudinal amplitude would have a pole at $q^2\to0$.
However, it should be noted that for the longitudinal transversity amplitude one should consider $A_0 \propto \sqrt{q^2} \, H_V(\lambda=0)$ and hence 
there is no pole at $q^2\to0$. 

\section{Additional fit results}\label{sec:additionalfits}

\begin{table}[h!]
\begin{center}
\setlength\extrarowheight{3pt}
\scalebox{0.8}{
\begin{tabular}{|c|c|}
\hline 
  \multicolumn{2}{|c|}{All observables with $\chi^2_{\rm SM}=118.8$} \\  [-3pt]
  \multicolumn{2}{|c|}{($\chi^2_{\rm min}=117.5;\; {\rm Pull}_{\rm SM}=0.3\sigma$)} \\ 
\hline \hline
\multicolumn{2}{|c|}{$\delta C_{10}$}\\ 
\multicolumn{2}{|c|}{$ 0.27 \pm 0.28 $} \\ 
\hline 
$C_{Q_{1}}$ & $C_{Q_{2}}$ \\
$-0.16 \pm 0.04$ & $0.11 \pm 0.27 $  \\
\hline 
\end{tabular}
}
\caption{Best fit values for the three operator $\{\delta C_{10}, C_{Q_{1}}$, $C_{Q_{2}}\}$ global fit to the $b \to s$ data, assuming 10\% error for the power corrections. 
\label{tab:allobs_3D_C10Q1Q2}} 
\end{center} 
\end{table}

\begin{table}[h!]
\begin{center}
\setlength\extrarowheight{3pt}
\scalebox{0.8}{
\begin{tabular}{|c|c|c|c|}
\hline 
  \multicolumn{4}{|c|}{All observables  with $\chi^2_{\rm SM}=118.8$} \\ 
  \multicolumn{4}{|c|}{$\chi^2_{\rm min}=101.1;\; {\rm Pull}_{\rm SM}=2.7\sigma$)} \\ 
\hline \hline
\multicolumn{2}{|c|}{$\delta C_{10}^{\mu}$} &  \multicolumn{2}{c|}{$\delta  C_{10}^{e}$}\\ 
\multicolumn{2}{|c|}{$  0.38 \pm 0.28 $} &  \multicolumn{2}{c|}{$-0.75 \pm 0.37$}\\ 
\hline 
$C_{Q_{1}}^{\mu}$ & $C_{Q_{2}}^{\mu}$ & $C_{Q_{1}}^{e}$ & $C_{Q_{2}}^{e}$ \\
$-0.10 \pm 0.31$ & $ 0.02 \pm 0.22 $ & $ 0.04 \pm 2.02 $ & $ 0.12 \pm 1.83 $ \\
\hline 
\end{tabular}
}
\caption{Best fit values for the six operator $\{\delta C_{10}^{e,\mu}, C_{Q_{1}}^{e,\mu}$, $C_{Q_{2}}^{e,\mu}\}$ global fit to the $b \to s$ data, assuming 10\% error for the power corrections.
\label{tab:allobs_6D_C10Q1Q2primes}} 
\end{center} 
\end{table}

\begin{table}[h!]
\begin{center}
\setlength\extrarowheight{3pt}
\scalebox{0.8}{
\begin{tabular}{|c|c|c|c|}
\hline 
  \multicolumn{4}{|c|}{All observables  with $\chi^2_{\rm SM}=118.8$} \\ 
  \multicolumn{4}{|c|}{($\chi^2_{\rm min}=81.2;\; {\rm Pull}_{\rm SM}=4.8\sigma$)} \\ 
\hline \hline
\multicolumn{2}{|c|}{$\delta C_7$} &  \multicolumn{2}{c|}{$\delta C_8$}\\ 
\multicolumn{2}{|c|}{$  0.02 \pm 0.05 $} &  \multicolumn{2}{c|}{$-0.08 \pm 0.70$}\\ 
\hline 
$\delta C_{9}^{\mu}$ & $\delta C_{9}^{e}$ & $\delta C_{10}^{\mu}$ & $\delta C_{10}^{e}$ \\
$-1.16 \pm 0.23$ & $ -2.38   \pm 2.23 $ & $ 0.00   \pm 0.23  $ & $  -1.38   \pm 0.53  $ \\
\hline 
\end{tabular}
}
\caption{Best fit values for the six operator $\{\delta C_7, \delta C_8, \delta C_{9}^{e,\mu}, \delta C_{10}^{e,\mu}\}$ global fit to the $b \to s$ data, assuming 10\% error for the power corrections.
\label{tab:allobs_6D_C78910}} 
\end{center} 
\end{table}

\begin{table}[h!]
\begin{center}
\setlength\extrarowheight{3pt}
\scalebox{0.8}{
\begin{tabular}{|c|c|c|c|}
\hline 
  \multicolumn{4}{|c|}{All observables  with $\chi^2_{\rm SM}=118.8$} \\ 
  \multicolumn{4}{|c|}{($\chi^2_{\rm min}=81.0;\; {\rm Pull}_{\rm SM}=4.1\,(4.5)\sigma$)} \\ 
\hline \hline
\multicolumn{2}{|c}{$\delta C_7$} &  \multicolumn{2}{|c|}{$\delta C_8$}\\ 
\multicolumn{2}{|c}{$ 0.02 \pm 0.05 $} &  \multicolumn{2}{|c|}{$ -0.07   \pm 0.69  $}\\ 
\hline \hline
$\delta C_{9}^{\mu}$  & $\delta C_{9}^{e}$      &   $\delta C_{10}^{\mu}$ & $\delta C_{10}^{e}$ \\
$ -1.17   \pm 0.23 $  & $ -2.38   \pm 2.24 $    &   $ -0.03 \pm 0.25  $    & $ -1.40   \pm 0.53  $ \\
\hline \hline
$C_{Q_{1}}^{\mu}$ & $C_{Q_{1}}^{e}$ & $C_{Q_{2}}^{\mu}$ & $C_{Q_{2}}^{e}$ \\ 
$ -0.14   \pm 0.15 $     & undetermined & $ 0.08 \pm 0.26 $  & undetermined \\  
\hline 
\end{tabular}
}
\caption{Best fit values for the ten operator $\{\delta C_7, \delta C_8, \delta C_{9}^{e,\mu}, \delta C_{10}^{e,\mu}, C_{Q_{1}}^{e,\mu}, C_{Q_{2}}^{e,\mu}\}$ global fit to the $b \to s$ data, assuming 10\% error for the power corrections.
\label{tab:allobs_10D_C78910C12}} 
\end{center} 
\end{table}

\clearpage

\providecommand{\href}[2]{#2}\begingroup\raggedright\endgroup

\end{document}